\RequirePackage{fix-cm}
\documentclass[natbib]{svjour3}
\smartqed  
\usepackage{etex}
\pdfoutput=1
\AtBeginDocument{

}

\usepackage{lscape}
\usepackage{comment}

\usepackage{graphicx}
\usepackage[utf8]{inputenc}
\usepackage[T1]{fontenc}
\usepackage{listings}
\usepackage{xcolor}
\usepackage{hyperref}
\usepackage{tabularx}
\usepackage{booktabs}
\usepackage{mdframed} 
\mdfsetup{skipabove=3pt,skipbelow=0pt}
\usepackage{verbatim} 
\usepackage{paralist}
\usepackage{listings}
\usepackage{courier}
\usepackage{xspace}
\usepackage{balance}
\usepackage{algorithm} 
\usepackage{algorithmic}
\usepackage{multirow}
\usepackage{amsmath}
\usepackage{amssymb}
\usepackage{pifont}
\newcommand{\cmark}{\text{\ding{51}}}
\newcommand{\xmark}{\text{\ding{55}}}

\newcolumntype{s}{>{\hsize=.75\hsize}X}
\newcolumntype{m}{>{\hsize=1.25\hsize}X}

\newcommand{\tabincell}[2]{\begin{tabular}{@{}#1@{}}#2\end{tabular}}

\usepackage{listings}
\usepackage{color}
\usepackage{enumitem} 
\definecolor{dkgreen}{rgb}{0,0.6,0}
\definecolor{gray}{rgb}{0.5,0.5,0.5}
\definecolor{mauve}{rgb}{0.58,0,0.82}

\begin{document}

\title{Alleviating Patch Overfitting with Automatic Test Generation: A Study of Feasibility and Effectiveness for the Nopol Repair System}
\author{Zhongxing Yu         \and
        Matias Martinez \and
        Benjamin Danglot \and
        Thomas Durieux \and
        Martin Monperrus
}

\institute{Zhongxing Yu . Benjamin Danglot . Thomas Durieux \at
              Inria Lille - Nord Europe, Avenue du Halley, 59650 Villeneuve-d’Ascq, France \\
              \email{zhongxing.yu@inria.fr . benjamin.danglot@inria.fr . thomas.durieux@inria.fr}           
           \and
           Matias Martinez \at
              University of Valenciennes, Malvache Building, Campus Mont Houy, 59313 Valenciennes Cedex 9, France \\
              \email{matias.martinez@univ-valenciennes.fr}           
              \and
           {Martin Monperrus} \at
              School of Computer Science and Communication, KTH Royal Institute of Technology, Stockholm, Sweden \\
              \email{martin.monperrus@csc.kth.se}           
}

\maketitle

\begin{abstract}
Among the many different kinds of program repair techniques, one widely studied family of techniques is called test suite based repair. However, test suites are in essence input-output specifications and are thus typically inadequate for completely specifying the expected behavior of the program under repair. Consequently, the patches generated by test suite based  repair techniques can just overfit to the used test suite, and fail to generalize to other tests. We deeply analyze the overfitting problem in program repair and give a classification of this problem. This classification will help the community to better understand and design techniques to defeat the overfitting problem. We further propose and evaluate an approach called UnsatGuided, which aims to alleviate the overfitting problem for synthesis-based repair techniques with automatic test case generation. The approach uses additional automatically generated tests to strengthen the repair constraint used by synthesis-based repair techniques. We analyze the effectiveness of UnsatGuided:
1) analytically with respect to alleviating two different kinds of overfitting issues;
2) empirically based on an experiment over the 224 bugs of the Defects4J repository.
The main result is that automatic test generation is effective in alleviating one kind of overfitting issue--regression introduction, but due to oracle problem, has minimal positive impact on alleviating the other kind of overfitting issue--incomplete fixing. 

\end{abstract}

\section{Introduction}

Automated program repair holds out the promise of saving debugging costs and patching buggy programs more quickly than humans. Given this great potential, there has been a surge of research on automated program repair in recent years and several different techniques have been proposed (\cite{genprog,semfix,nopol,tsepei,long2017automatic}). These techniques differ in various ways, such as the kinds of used oracles and the fault classes they target\footnote{In this paper, we use ``fault'' and ``bug'' interchangeably.} (\cite{Monperrus2015}). 

Among the many different techniques proposed, one widely studied family of techniques is called test suite based repair. Test suite based repair starts with some passing tests as the specification of the expected behavior of the program and at least one failing test as a specification of the bug to be repaired, and aims at generating patches that make all the tests pass. Depending the patch generation strategy, test suite based repair can further be informally divided into two general categories: generate-and-validate techniques and synthesis-based techniques. Generate-and-validate techniques use certain methods such as genetic programming to first generate a set of candidate patches, and then validate the generated patches against the test suite. Representative examples in this category include GenProg (\cite{genprog}), PAR (\cite{kim2013automatic}) and SPR (\cite{spr}). Synthesis-based techniques first use test execution information to build a repair constraint, and then use a constraint solver to synthesize a patch. Typical examples in this category include SemFix (\cite{semfix}), Nopol (\cite{nopol}), and Angelix (\cite{Mechtaev:2016:ASM:2884781.2884807}). Empirical studies have shown the promise of test suite based repair techniques in tackling real-life bugs in real-life systems. For instance, GenProg (\cite{genprog}) and Angelix (\cite{Mechtaev:2016:ASM:2884781.2884807}) can generate repairs for large-scale real-world C programs, while ASTOR (\cite{astor2016}) and Nopol (\cite{nopol}) have given encouraging results (\cite{defects4j-repair}) on a set of real-life Java programs from the Defects4j benchmark (\cite{JustJE2014}).

However, test suites are in essence input-output specifications and are therefore typically inadequate for completely specifying the expected behavior.
Consequently, the patches generated by test suite based program repair techniques pass the test suite, yet may be incorrect. The patches that are overly specific to the used test suite and fail to generalize to other tests are called overfitting patches (\cite{smith2015cure}). Overfitting indeed threats the validity of test suite based repair techniques and some recent studies have shown that a significant portion of the patches generated by test suite based repair techniques are overfitting patches (\cite{smith2015cure,qi2015efficient,defects4j-repair,leoverfitting}). 

In this paper, we deeply analyze the overfitting problem in program repair and identify two kinds of overfitting issues: incomplete fixing and regression introduction. Our empirical evaluation shows that both kinds of overfitting issues are common. Based on the overfitting issues that an overfitting patch has, we further define three kinds of overfitting patches. This characterization of overfitting will help the community to better understand the overfitting problem in program repair, and will hopefully guide the development of techniques for alleviating overfitting. 

We further propose an approach called UnsatGuided, which aims to alleviate the overfitting problem for synthesis-based techniques. Given the recent significant progress in the area of automatic test generation, UnsatGuided makes use of automatic test case generation technique to obtain additional tests and then integrate the automatically generated tests into the synthesis process. The intuition behind UnsatGuided is that additional automatically generated tests can supplement the manually written tests to strengthen the repair constraint, and synthesis-based techniques can thus use the strengthened repair constraint to synthesize patches that suffer less from overfitting. To generate tests that can detect problems besides \textit{crashes} and \textit{uncaught exceptions}, state-of-art automatic test generation techniques generate tests that include assertions encoding the behavior observed during test execution on the current program. By using such automatic test generation techniques on the program to be repaired, some of the generated tests can possibly assert buggy behaviors and these tests with wrong oracles can mislead the synthesis process. UnsatGuided tries to identify and discard tests with likely wrong oracles through the idea that if the additional repair constraint from a generated test has a contradiction with the repair constraint established using the manually written test suite, then the generated test is likely to be a test with wrong oracle. 

We analyze the effectiveness of UnsatGuided with respect to alleviating different kinds of overfitting issues. We then set up an empirical evaluation of UnsatGuided, which uses Nopol (\cite{nopol}) as the synthesis-based technique and EvoSuite (\cite{ESECFSE11}) as the automatic test case generation technique. The evaluation uses 224 bugs of the Defects4J repository (\cite{JustJE2014}) as benchmark. The results confirm our analysis and show that UnsatGuided 1) is effective in alleviating overfitting issue of regression introduction for 16/19 bugs; 2) does not break already correct patches; 3) can help a synthesis-based repair technique to generate additional correct patches. 

To sum up, the contributions of this paper are: 

\begin{itemize}
\item An analysis of the overfitting problem in automated program repair and a classification of overfitting.

\item An approach, called UnsatGuided, to alleviate the overfitting problem for synthesis-based repair techniques.

\item An analysis of the effectiveness of UnsatGuided in alleviating different kinds of overfitting issues, and the identification of deep limitations of using automatic test case generation to alleviate overfitting.

\item An empirical evaluation of the prevalence of different kinds of overfitting issues on 224 bugs of the Defects4J repository, as well as an extensive evaluation of the effectiveness of UnsatGuided in alleviating the overfitting problem. 

\end{itemize}

The remainder of this paper is structured as follows. We first present related work in Section 2. Section 3 first provides our analysis of the overfitting problem and the classification of overfitting issues and overfitting patches, then gives the algorithm of the proposed approach UnsatGuided, and finally analyzes the effectiveness of UnsatGuided. Section 4 presents an empirical evaluation of the prevalence of different kinds of overfitting issues and the effectiveness of UnsatGuided, followed by Section 5 which concludes this paper.
This paper is a major revision of an Arxiv preprint (\cite{Yu2017Test4Repair}).

\section{Related Work}

\subsection{Program Repair}
Due to the high cost of fixing bugs manually, there has been a surge of research on automated program repair in recent years. Automated program repair aims to correct software defects without the intervention of human developers, and many different kinds of techniques have been proposed recently. For a complete picture of the field, readers can refer to the survey paper (\cite{Monperrus2015}). Generally speaking, automated program repair involves two steps. To begin with, it analyzes the buggy program and uses techniques such as genetic programming (\cite{genprog}), program synthesis (\cite{semfix}) and machine learning (\cite{prophet}) to produce one or more candidate patches. Afterwards, it validates the produced candidate patches with an oracle that encodes the expected behavior of the buggy program. Typically used oracles include test suites (\cite{genprog, semfix}), pre- and post-conditions (\cite{Weitse}), and runtime assertions (\cite{Perkins2009}). The proposed automatic program repair techniques can target different kinds of faults. While some automatic program techniques target the general types of faults and do not require the fault types to be known in advance, a number of other techniques can only be applied to specific types of faults, such as null pointer exception (\cite{SANER2017}), integer overflow (\cite{Brumley07}), buffer overflow (\cite{dsn2014}), memory leak (\cite{memoryfixing}), and error handling bugs (\cite{errorhandlingFSE}).

\subsection{Test Suite Based Program Repair}
Among the various kinds of program repair techniques proposed, a most widely studied and arguably the standard family of techniques is called test suite based repair. The inputs to test suite based repair techniques are the buggy program and a test suite, which contains some passing tests as the specification of the expected behavior of the program and at least one failing test as a specification of the bug to be repaired. The output is one or more candidate patches that make all the test cases pass. Typically, test suite based repair techniques first use some fault localization techniques (\cite{tarantula,sober,yzxicse,YuIST,yuhase,predicateswitching}) to identify the most suspicious program statements. Then, test suite based repair techniques use some patch generation strategies to patch the identified suspicious statements. Based on the used patch generation strategy, test suite based repair techniques can further be divided into generate-and-validate techniques and synthesis-based techniques. 

Generate-and-validate repair techniques first search within a search space to generate a set of patches, and then validate them against the test suite. GenProg (\cite{genprog}), one of the earliest generate-and-validate techniques, uses genetic programming to search the repair space and generates patches that consist of code snippets copied from elsewhere in the same program. PAR (\cite{kim2013automatic}) shares the same search strategy with GenProg but uses 10 specialized patch templates derived from human-written patches to construct the search space. RSRepair (\cite{rsrepair}) has the same search space as GenProg but uses random search instead, and the empirical evaluation shows that random search can be as effective as genetic programming. AE (\cite{6693094}) employs a novel deterministic search strategy and uses program equivalence relation to reduce the patch search space. SPR (\cite{spr}) uses a set of predefined transformation schemas to construct the search space, and patches are generated by instantiating the schemas with condition synthesis techniques. Prophet (\cite{prophet}) applies probabilistic models of correct code learned from successful human patches to prioritize candidate patches so that the correct patches could have higher rankings. Given that most of the proposed repair systems target only C code, jGenProg, as implemented in ASTOR (\cite{astor2016}), is an implementation of GenProg for Java code. 

Synthesis-based techniques first use the input test suite to extract a repair constraint, and then leverage program synthesis to solve the constraint and get a patch. The patches generated by synthesis-based techniques are generally by design correct with respect to the input test suite. 
SemFix (\cite{semfix}), the pioneer work in this category of repair techniques, performs controlled symbolic execution on the input tests to get symbolic constraints, and uses code synthesis to identify a code change that makes all tests pass. The target repair locations of SemFix are assignments and boolean conditions. To make the generated patches more readable and comprehensible for human beings, DirectFix (\cite{directfix}) encodes the repair problem into a partial Maximum Satisfiability problem (MaxSAT) and uses a suitably modified Satisfiability Modulo Theory (SMT) solver to get the solution, which is finally converted into the concise patch. Angelix (\cite{Mechtaev:2016:ASM:2884781.2884807}) uses a lightweight repair constraint representation called “angelic forest” to increase the scalability of DirectFix. Nopol (\cite{nopol}) uses multiple instrumented test suite executions to synthesize a repair constraint, which is then transformed into a SMT problem and a feasible solution to the problem is finally returned as a patch. Nopol addresses the repair of buggy \emph{if} conditions and missing preconditions. S3 (\cite{S3FSE}) aims to synthesize more generalizable patches by using three components: a domain-specific language (DSL) to customize and constrain search space, an enumeration-based search strategy to search the space, and finally a ranking function to rank patches. 

While test suite based repair techniques are promising, an inherent limitation of them is that the correctness specifications used by them are the test suites, which are generally available but rarely exhaustive in practice. As a result, the generated patches may just overfit to the available tests, meaning that they will break untested but desired functionality. Several recent studies have shown that overfitting is a serious issue associated with test suite based repair techniques. Qi et al. (\cite{qi2015efficient}) find that the vast majority of patches produced by GenProg, RSRepair, and AE avoid bugs simply by functionality deletion. A subsequent study by Smith et al. (\cite{smith2015cure}) further confirms that the patches generated by GenProg and RSRepair fail to generalize. The empirical study conducted by Martinez et al. (\cite{defects4j-repair}) reveals that among the 47 bugs fixed by jGenProg, jKali, and Nopol, only 9 bugs are correctly fixed. More recently, the study by Le et al. (\cite{leoverfitting}) again confirms the severity of the overfitting issue for synthesis-based repair techniques. Moreover, the study also investigates how test suite size and provenance, number of failing tests, and semantics-specific tool settings can affect overfitting issues for synthesis-based repair techniques. Given the seriousness and importance of the overfitting problem, Yi et al. (\cite{correlationstudy}) explore the correlation between test suite metrics and the quality of patches generated by automated program repair tetchiness, and they find that with the increase of traditional test suite metrics, the quality of the generated patches also tend to improve.

To gain a better understanding of the overfitting problem in program repair, we conduct a deep analysis of it and give the classification of overfitting issues and overfitting patches. We wish the classifications can facilitate future work on alleviating the overfitting problem in program repair. In addition, given the recent progress in the area of automatic test generation, we investigate the feasibility of augmenting the initial test suite with additional automatically generated tests to alleviate the overfitting problem. More specifically, we propose an approach called UnsatGuided, which aims to alleviate the overfitting problem for synthesis-based repair techniques. The effectiveness of UnsatGuided for alleviating different kinds of overfitting issues is analyzed and empirically verified, and we also point out the deep limitations of using automatic test generation to alleviate overfitting. 

In the literature, there are several works that try to use test case generation to alleviate the overfitting problem in program repair. Xin and Reiss (\cite{qixinISSTA}) propose an approach to identify overfitting patches through test case generation, which generates new test inputs that focus on the semantic differences brought by the patches and relies on human beings to add oracles for the inputs. Yang et al. (\cite{yangFSE}) aim to filter overfitting patches for generate-and-validate repair techniques through a framework named \emph{Opad}, which uses fuzz testing to generate tests and relies on two inherent oracles, crash and memory-safety, to enhance validity checking of generated patches. By heuristically comparing the similarity of different execution traces, Liu et al. (\cite{liu2017}) also aim to identify overfitting patches generated by test suite based repair techniques. UnsatGuided is different from these works. On the one hand, these three works all try to use generated tests to identify overfitting patches generated by test suite based repair techniques and the generated tests are not used by the run of the repair algorithm itself. However, our aim is to improve the patch generated using manually written test suite and the generated tests are used by the repair algorithm to supplement the manually written test suite so that a better repair specification can be obtained. On the other hand, our work does not assume the specificity of the used oracle while the work by Xin and Reiss (\cite{qixinISSTA}) uses the human oracle and the work by Yang et al. (\cite{yangFSE}) uses the crash and memory-safety oracles. 

\subsection {Automatic Test Case Generation}

Despite tests are often created manually in practice, much research effort has been put on automated test generation techniques. In particular, a number of automatic test generation tools for mainstream programming languages have been developed over the past few years. These tools typically rely on techniques such as random test generation, search-based test generation and dynamic symbolic execution. 

For Java, Randoop (\cite{randoop}) is the well-known random unit test generation tool. Randoop uses feedback-directed random testing to generate unit tests, and it works by iteratively extending method call sequences with randomly selected method calls and randomly selected arguments from previously constructed sequences. As Randoop test generation process uses a bottom-up approach, it cannot generate tests for a specific class. Other random unit test generation tools for Java include JCrasher (\cite{jcrasher}), CarFast (\cite{carfast}), T3 (\cite{Prasetya2014}), TestFul (\cite{testful}) and eToc (\cite{Tonella:2004:ETC:1013886.1007528}). There are also techniques that use various kinds of symbolic execution, such as symbolic PathFinder (\cite{Pasareanu:2010:SPS:1858996.1859035}) and DSC (\cite{Islam:2010:DTC:1868321.1868326}). EvoSuite (\cite{ESECFSE11}) is the state-of-art search-based unit test generation tool for Java and can target a specific class. It uses an evolutionary approach to derive test suites that maximize code coverage, and generates assertions that encode the current behavior of the program. 

In the C realm, DART (\cite{godefroid2005dart}), CUTE (\cite{sen2005cute}), and KLEE (\cite{cadar2008klee}) are three representatives of automatic test case generation tools for C. Symbolic execution is used in conjunction with concrete execution by these tools to maximize code coverage. In addition, Pex (\cite{Tillmann:2008:PWB:1792786.1792798}) is a popular unit test generation tool for C\# code based on dynamic symbolic execution.

\section{Analysis and Alleviation of the Overfitting Problem}
\label{sec:analysis}

In this section, we first introduce a novel classification of overfitting issues and overfitting patches. Then, we propose an approach called UnsatGuided for alleviating the overfitting problem for synthesis-based repair techniques. We finally analyze the effectiveness of UnsatGuided with respect to different overfitting kinds and point out the profound limitation of using automatic test generation to alleviate overfitting. 

\subsection{Core Definitions} 

Let us reason about the input space $I$ of a program $P$.
We consider modern object-oriented programs, where an input point is composed of one or more objects, interacting through a sequence of methods calls. 
In a typical repair scenario, the program is almost correct and thus
a bug only affects the program behavior of a portion of the input domain, which we call the ``buggy input domain'' $I_{bug}$.
We call the rest of the input domain, for which the program behaviors are considered correct as $I_{correct}$. 
By definition, a patch generated by an automatic program repair technique has an impact on program behaviors, i.e., it changes the behaviors of a portion of the input domain. We use $I_{patch}$ to denote this input domain which is impacted by a patch. For input points within $I_{bug}$ whose behaviors have been changed by a patch, the patch can either correctly or incorrectly change the original buggy behaviors. We use $I_{patch=}$ to denote the input points within $I_{bug}$ whose behaviors have been incorrectly changed by a patch, i.e., the newly behaviors of these input points brought by the patch are still incorrect. Meanwhile, we use $I_{patch{\cmark}}$ to denote the input points within $I_{bug}$ whose behaviors have been correctly changed by a patch. If the patch involves changes to behaviors of input points within $I_{correct}$, then the original correct behaviors of these input points will undesirably become incorrect and we use $I_{patch{\xmark}}$ to denote these input points within $I_{correct}$ broken by the patch. Obviously, the union of $I_{patch=}$, $I_{patch{\cmark}}$ and $I_{patch{\xmark}}$ makes up $I_{patch}$.

For simplicity, hereafter when we say some input points within $I_{bug}$ are repaired by a patch, we mean the original buggy behaviors of these input points have been correctly changed by the patch. Similarly, when we say some input points within $I_{correct}$ are broken by a patch, we mean the original correct behaviors of these input points have been incorrectly changed by the patch. 

Note as a patch generated by test suite based program repair techniques, the patch will at least repair the input points corresponding to the original failing tests. In other words, the intersection of $I_{patch{\cmark}}$ and $I_{bug}$ will always not be empty ($I_{patch{\cmark}} \cap I_{correct} \neq \emptyset $).

\subsection{Classification of Overfitting} 
For a given bug, a perfect patch repairs all input points within $I_{bug}$ and does not break any input points within $I_{correct}$.
However, due to the incompleteness of the test suite used to drive the repair process, the generated patch may not be ideal and just overfit to the used tests.
Depending on how a generated patch performs with respect to the input domain $I_{bug}$ and $I_{correct}$, we define two kinds of overfitting issues, which are consistent with the problems for human patches introduced by Gu et al (\cite{Guicse}).

\vspace{1.5mm}
\textbf{Incomplete fixing}: Some but not all input points within $I_{bug}$ are repaired by the generated patch. In other words, $I_{patch{\cmark}}$ is a proper subset of $I_{bug}$ ($I_{patch{\cmark}} \subset I_{bug}$).

\vspace{1.5mm} 
\textbf{Regression introduction}: Some input points within $I_{correct}$ are broken by the generated patch. In other words, $I_{patch{\xmark}}$ is not an empty set ($I_{patch{\xmark}} \neq \emptyset $).

\vspace{1.5mm} 

Based on these two different kinds of overfitting issues, we further define three different kinds of overfitting patches.

\begin{figure}
\centering
 \includegraphics[scale=0.389]{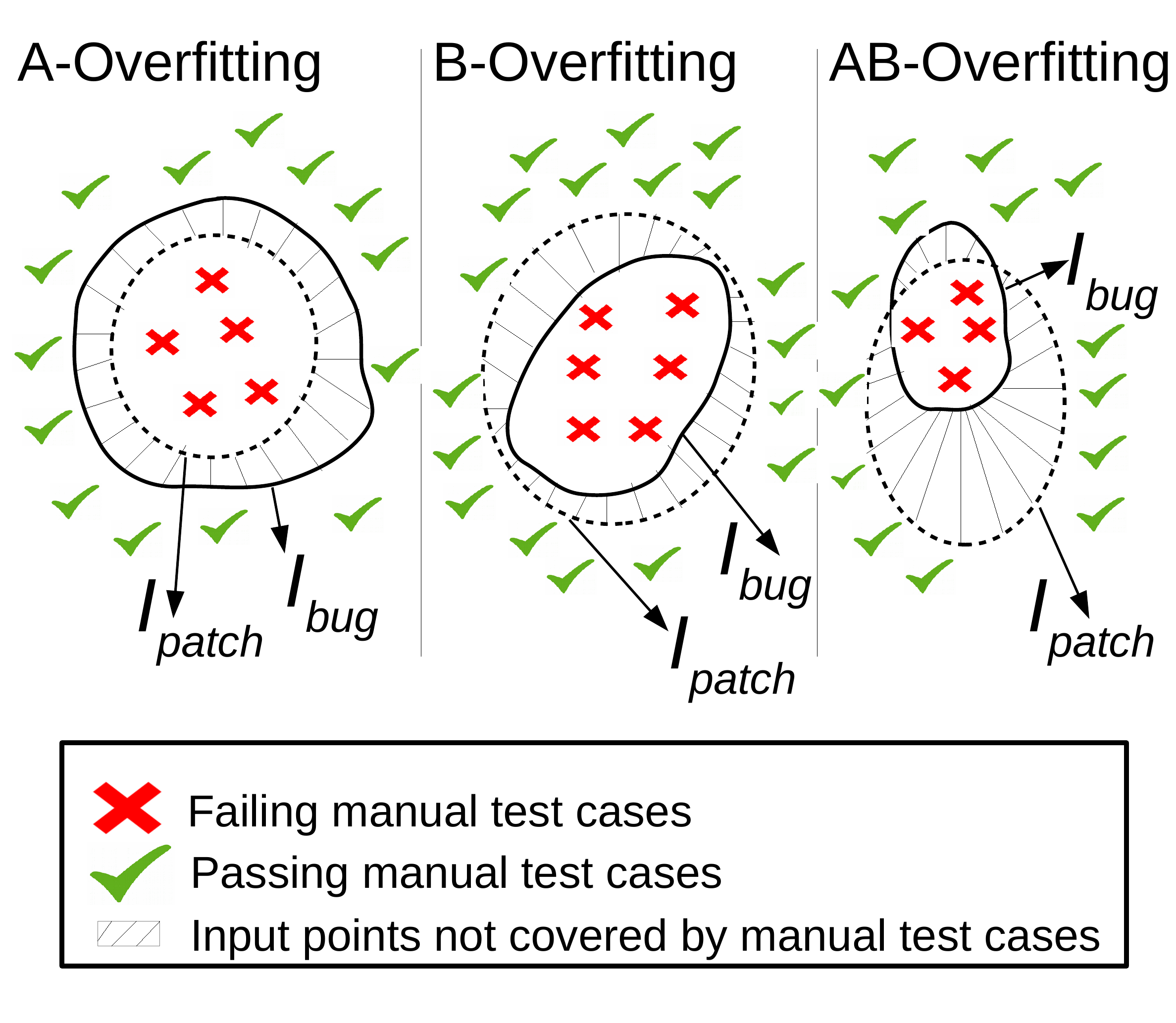}
 \caption{
A-Overfitting patch is a partial patch on a portion of the buggy input domain.
B-Overfitting patch breaks correct behaviors outside the buggy input domain.
AB-Overfitting patch partially fixes the buggy input domain and also breaks some correct behaviours.
}
 \label{fig:ibug_overfitting}
\end{figure}

\vspace{1.5mm} 
\textbf{A-Overfitting patch}: The overfitting patch only has the overfitting issue of incomplete fixing ($I_{patch{\cmark}} \subset I_{bug} \land I_{patch{\xmark}} = \emptyset$). This kind of overfitting patch can be considered as a ``partial patch''. It encompasses the worst case where there is one single failing test and the overfitting patch fixes the bug only for the input point specified in this specific failing test.

\vspace{1.5mm} 
\textbf{B-Overfitting patch}: The overfitting patch only has the overfitting issue of regression introduction ($I_{patch{\cmark}} = I_{bug} \land I_{patch{\xmark}} \neq \emptyset$). Note that this kind of overfitting patch correctly repairs all input points within the buggy input domain $I_{bug}$ but at the same time breaks some already correct behaviors of the buggy program under repair.

\vspace{1.5mm} 
\textbf{AB-Overfitting patch}: The overfitting patch has both overfitting issues of incomplete fixing and regression introduction at the same time ($I_{patch{\cmark}} \subset I_{bug} \land I_{patch{\xmark}} \neq \emptyset$). This kind of overfitting patch correctly repairs some but not all input points within the buggy input domain $I_{bug}$ and also introduces some regressions. 

\vspace{1.5mm} 
\autoref{fig:ibug_overfitting} gives an illustration of these three different kinds of overfitting patches. This characterization of overfitting in program repair is independent of the technique presented in this paper and can be used by the community to better design techniques to defeat the overfitting problem.

\subsection{UnsatGuided: Alleviating the Overfitting Problem for Synthesis-based Repair Techniques}

In this section, we propose an approach called UnsatGuided, which aims to alleviate the overfitting problem for synthesis-based repair techniques. The approach aims to strengthen the correctness specification so that the resulting generated patches are more likely to generalize over the whole input domain.
It achieves the aim by using additional tests generated by an automatic test case generation technique. 
We first give some background knowledge about automatic test case generation techniques and then give the details of the proposed approach. 

\subsubsection{The Bug-exposing Test Problem}

In the context of regression testing, automatic test case generation techniques typically use the current behavior of the program itself as the oracle (\cite{randoop,Xietaoecoop})\footnote{We do no uses the techniques that generate assertions from runs of different program versions (\cite{TaoDiffGenAR,EvansFSE}).}. 
We consider those typical regression test generation techniques in this paper and denote an arbitrary technique as $T_{reg}$. 

For a certain buggy version, $T_{reg}$ may generate both input points within the buggy input domain $I_{bug}$ and the correct input domain $I_{correct}$. For instance, suppose we have a calculator which incorrectly implements the \texttt{add} function for achieving the addition of two integers. The code is buggy on the input domain $(10, \_)$ (where \_ means any integer except 0) and is implemented as follows:

\begin{lstlisting}
add(x,y) {
  if (x == 10) return x-y;
  else return x+y; 
}
\end{lstlisting}

First, assume that $T_{reg}$ generates a test in the correct input domain $I_{correct}$, say for input point $(5,5)$.
The resulting test, which uses the existing behavior as oracle, will be \texttt{assertEquals(10, add(5,5))}. 
Then consider what happens when the generated test lies in $I_{bug}$, say for input point $(10,8)$. In this case, $T_{reg}$ would generate the test \texttt{assertEquals(2, add(10,8))}.

If the input point of a generated test lies in $I_{bug}$, the synthesized assertion will assert the presence of the actual buggy behavior of the program under test, i.e., the generated assertion encodes the buggy behavior. 
In such a case, if the input point of a generated test lies in $I_{bug}$, it is called a ``bug-exposing test'' in this paper. 
Otherwise, the test is called a ``normal test'' if its input point lies in $I_{correct}$. 

In the context of test suite based program repair, the existence of bug-exposing tests is a big problem. Basically, if a repair technique finds a patch that satisfies bug-exposing tests, then the buggy behavior is kept. In other words, it means that some of the generated tests can possibly enforce bad behaviors related with the bug to be repaired.

\subsubsection{UnsatGuided: Incremental Test Suite Augmentation for Alleviating the Overfitting Problem for Synthesis-based Repair Techniques}
\label{sec:example}

The overfitting problem for synthesis-based repair techniques such as SemFix and Nopol arises because the repair constraint established using an incomplete test suite is not strong enough to fully express the intended semantics of a program.
Our idea is to strengthen the initial repair constraint by augmenting the initial test suite with additional automatically generated tests. We wish that a stronger repair constraint would guide synthesis-based repair techniques towards better patches, i.e., patches that are correct or at least suffer less from overfitting. 

The core problem to handle is the possible existence of bug-exposing test(s) among the tests generated by an automatic test case generation technique. We cannot directly supply all of the generated tests to a synthesis-based repair technique because bug-exposing tests can mislead the synthesis repair process and force incorrect behaviors to be synthesized. 
 
To handle this core conceptual problem, we now present an approach called UnsatGuided, which gradually makes use of the new information provided by each automatically generated test to build a possibly stronger final repair constraint. The key underlying idea is that if the additional repair constraint enforced by an automatically generated test has logical contradictions with the repair constraint established so far, then the generated test is likely to be a bug-exposing test and is discarded. 

\paragraph{Example}
To help understanding, we use the following toy program to illustrate it. The inputs are any integers and there is an error in the condition which results in buggy input domain $I_{bug}$ = \{5, 6, 7\}.
Suppose we use component based repair synthesis (\cite{JhaICSE}) to synthesize the correct condition, and to make the explanation easy, we further assume the available components include only variable x, the relational operators < (less than) and > (greater than), logical operator \&\& (logical and), and finally any integer constants. For the three buggy inputs, regression test generation technique $T_{reg}$ considered in this paper can generate bug-exposing tests \texttt{assertEquals(4, f(5))}, \texttt{assertEquals(5, f(6))}, and \texttt{assertEquals(6, f(7))}. Each test is of the form \texttt{assertEquals(${O}$,${f(I)}$)}, which specifies that the expected return value of the program is ${O}$ when the input is ${I}$. For other input points, manually written tests and tests generated by $T_{reg}$ are the same. Each test \texttt{assertEquals(${O}$,${f(I)}$)} will impose a repair constraint of the form ${x}$=${I}$$\rightarrow$${f(I)=O}$. The repair constraint imposed by a set of tests $\{t_{i}|$\texttt{assertEquals($O_{i},f(I_{i})$)},$1\leqslant i\leqslant N\}$ will be $\bigwedge\limits_{i=1}^N$ (${x}$=$I_{i}$$\rightarrow$${f(I_{i})=O_{i}}$). The repair constraint and available components are then typically encoded into a SMT problem, and a satisfying SMT model is then translated back into a synthesized expression which provably satisfies the repair constraint imposed by the tests. To achieve the encoding, techniques such as concrete execution (\cite{nopol}) and symbolic execution (\cite{semfix}) can be used. 

\begin{lstlisting}
int f(int x)  {
  if (x>0&&x<5) //faulty, correct condition should be (x>0&&x<8)
      x++;
  else
      x--;
  return x;
}
\end{lstlisting}

For this example, suppose the manually written tests \texttt{assertEquals(-1, f(0))}, \texttt{assertEquals(2, f(1))}, \texttt{assertEquals(8, f(7))}, and \texttt{assert Equals(9, f(10))} are provided initially. Using the repair constraint $( x = 0 $ $\rightarrow f \left( 0 \right) = - 1) \wedge \left( x = 1 \rightarrow f  \left( 1 \right) = 2 \right) \wedge \left( x = 7 \rightarrow f \left( 7 \right) = 8 \right) \wedge \left( x = 10 \rightarrow f \left( 10 \right) = 9 \right)$ enforced by these tests, the synthesis process can possibly synthesize a condition if (x>0 \&\& x<10), which is not completely correct as the repair constraint enforced by the 4 manual tests is not strong enough. 
If a bug-exposing test such as \texttt{assertEquals(4,f(5))} is generated by $T_{reg}$ and the repair constraint ($x=5\rightarrow f(5)=4$) imposed by it is added, the synthesis process cannot synthesize a condition as there is a contradiction between the repair constraint imposed by it and that imposed by the 4 manual tests. The contradiction happens because according to the the repair constraint imposed by manual tests and the available components used for synthesis, the calculation of any integer input between 1 and 7 should follow the same branch as integer inputs 1 and 7, consequently the return value should be 6 (not 4) when the integer input is 5. The core idea of UnsatGuided is to detect those contradictions and discard the bug exposing tests such as \texttt{assertEquals(4, f(5))}.

However, if a normal test such as \texttt{assertEquals(7, f(8))} is generated by $T_{reg}$ and the repair constraint ($x=8\rightarrow f(8)=7$) imposed by it is added, there is no contradiction and a stronger repair constraint can be obtained, which will enable the synthesis process to synthesize the correct condition if (x>0 \&\& x<8) in this specific example.  The core idea of UnsatGuided is to keep those valuable new tests for synthesizing and validating patches.

\paragraph{Algorithm}

\begin{algorithm}[t]
\begin{algorithmic}[1]
\REQUIRE{A buggy program $P$ and its manually written test suite $TS$}
\REQUIRE{A synthesis-based repair technique $T_{synthesis}$ and the time budget $TB$}
\REQUIRE{An automatic test case generation tool $T_{auto}$}
\ENSURE{A patch $pt$ to the buggy program $P$}
\STATE{$pt_{initial} \leftarrow T_{synthesis}(P, TS, TB)$}
\IF{$pt_{initial} = null$}
    \STATE{$pt \leftarrow null$}
\ELSE
    \STATE{$AGTS \leftarrow \emptyset$}
    \STATE{$pt \leftarrow pt_{initial}$}
    \STATE{$TS_{aug} \leftarrow TS$}
    \STATE{$t_{initial} \leftarrow getPatchGenTime(T_{synthesis}(P, TS, TB))$}
    \STATE{$\{file_{i}\}(i=1,2,...,n) \leftarrow getInvolvedFiles(pt_{initial})$ }
    \FOR{$i=1$ to $n$}
        \STATE{$AGTS \leftarrow AGTS \cup T_{auto}(P, file_{i})$}
    \ENDFOR
    \FOR{$j=1$ to $|AGTS|$}
        \STATE{$t_j \leftarrow AGTS(j)$}
        \STATE{$TS_{aug} \leftarrow TS_{aug} \cup \{t_j\}$}
        \STATE{$pt_{intern} \leftarrow T_{synthesis}(P, TS_{aug}, t_{initial} \times 2)$}
        \IF{$pt_{intern} \neq null$}
            \STATE{$pt \leftarrow pt_{intern}$}
        \ELSE
            \STATE{$TS_{aug} \leftarrow TS_{aug} - \{t_j\} $}
        \ENDIF
    \ENDFOR
\ENDIF
\RETURN{$pt$}
\end{algorithmic}
\caption{: Algorithm for the Proposed Approach UnsatGuided}
\label{alg:2}
\end{algorithm}

Algorithm \autoref{alg:2} describes the approach in detail. The algorithm takes as input a buggy program \emph{P} to be repaired, a manually written test suite \emph{TS} which contains some passing tests and at least one failing test, a synthesis-based repair technique $T_{synthesis}$, a time budget \emph{TB} allocated for the execution of $T_{synthesis}$, and finally an automatic test case generation tool $T_{auto}$ which uses a certain kind of automatic test case generation technique $T_{reg}$. The output of the algorithm is a patch \emph{pt} to the buggy program \emph{P}. 

The algorithm directly returns an empty patch if $T_{synthesis}$ generates no patches within the time budget (lines 2-3). In case $T_{synthesis}$ generates an initial patch $pt_{initial}$ within the time budget, the algorithm first conducts a set of initialization steps as follows: it sets the automatically generated test suite \emph{AGTS} to be an empty set (line 5), sets the returned patch \emph{pt} to be the initial patch $pt_{initial}$ (line 6), sets the augmented test suite $TS_{aug}$ to be the manually written test suite \emph{TS} (line 7), and gets the time used by $T_{synthesis}$ to generate the initial patch $pt_{initial}$ and sets $t_{initial}$ to be the value (line 8). Algorithm \autoref{alg:2} then identifies the set of files \{$file_i$\}(\emph{i}=1, 2,..., \emph{n}) involved in the initial patch $pt_{initial}$ (line 9) and for each identified file, it uses the automatic test case generation tool $T_{auto}$ to generate a set of tests that target behaviors related with the file and adds the generated tests to the automatically generated test suite \emph{AGTS} (lines 10-12). 

Next, the algorithm will use the test suite \emph{AGTS} to refine the initial patch $pt_{initial}$. For each test $t_j$ in the test suite \emph{AGTS} (line 14), the algorithm first adds it to the augmented test suite $TS_{aug}$ (line 15) and runs technique $T_{synthesis}$ with test suite $TS_{aug}$ and new time budget $t_{initial} \times 2$ against program \emph{P} (line 16). The new time budget is used to quickly identify tests that can potentially contribute to strengthening the repair constraint, and thus improve the scalability of the approach. Then, if the generated patch $pt_{intern}$ is not an empty patch, the algorithm updates the returned patch \emph{pt} with $pt_{intern}$ (lines 17-18). In other words, the algorithm deems test $t_j$ as a good test that can help improve the repair constraint. Otherwise, test $t_j$ is removed from the augmented test suite $TS_{aug}$ (lines 19-20) as $t_j$ is either a bug-exposing test or it slows down the repair process too much. After the above process has been completed for each test in the test suite \emph{AGTS}, the algorithm finally returns patch \emph{pt} as the desirable patch (line 24). 

\emph{Remark}: Note for a certain synthesis-based repair technique $T_{synthesis}$ that is used as the input, UnsatGuided does not make any changes to the patch synthesis process of $T_{synthesis}$ itself. In particular, most current synthesis-based repair techniques use component based synthesis to synthesize the patch, including Nopol (\cite{nopol}), SemFix (\cite{semfix}), Angelix (\cite{Mechtaev:2016:ASM:2884781.2884807}). For component-based synthesis, one important problem is selecting and using the build components. UnsatGuided keeps the original component selection and use strategy implemented by each synthesis-based repair technique. 

In addition, the order of trying each test in the test suite \emph{AGTS} matters. Once a test is deemed as helpful, it is added to the augmented test suite $TS_{aug}$ permanently and may impact the result of subsequent runs of other tests. The algorithm currently first uses the size of the identified files involved in the initial patch to determine the test generation order. The larger the size of an identified file, the earlier the test generation tool $T_{auto}$ will generate tests for it. We first generate tests for big files as big files, in general, encode more logic compared to small files, thus tests generated for them are more important. Then, the algorithm uses the creation time of generated test files and the order of tests in a generated test file to prioritize tests. The earlier a test file is created, the earlier its test(s) will be tried by the algorithm. And if a test file contains multiple tests, the earlier a test appears in the file, the earlier the algorithm will try it. Future work will prioritize generated tests according to their potential to improve the repair constraint. 

\subsection{Analysis of UnsatGuided}
\label{sec:analysis-unsatguided}

UnsatGuided uses additional automatically generated tests to alleviate the overfitting problem for synthesis-based repair techniques. The performance of UnsatGuided is mainly affected by two aspects. On the one hand, it is affected by how the synthesis-based repair techniques perform with respect to the original manually written test suite, i.e., it depends on the overfitting type of the original patch. On the other hand, it is affected by whether or not the automatic test case generation technique generates bug-exposing tests. Let us dwell on this. 

For ease of presentation, the initial repair constraint enforced by the manually written test suite is referred to as $RC_{initial}$, and the repair constraints enforced by the normal and bug-exposing tests generated by an automatic test case generation technique are referred to as $RC_{normal}$ and $RC_{buggy}$ respectively. Note due to the nature of test generation technique $T_{reg}$, $RC_{buggy}$ is wrong. Also, we use $P_{original}$ to denote the original patch generated using the manually written test suite by a synthesis-based repair technique. Finally, we also use the example program in \autoref{sec:example} to illustrate the key points of our analysis.

(1) \textbf{$P_{original}$ is correct}. In this case, $RC_{initial}$ is in general strong enough to drive the synthesis-based repair techniques to synthesize a correct patch. If the automatic test generation technique $T_{reg}$ generates bug-exposing tests, $RC_{buggy}$ will have contradictions with $RC_{initial}$ (note $RC_{buggy}$ is wrong) and UnsatGuided will recognize and discard these bug-exposing tests. Meanwhile, $RC_{normal}$ is likely to be already covered by $RC_{initial}$ and is not likely to make $P_{original}$ become incorrect by definition. It can happen that the synthesis process coincidentally synthesizes a correct patch even though $RC_{initial}$ is weak, but this case is relatively rare. Thus, UnsatGuided generally will not change an already correct patch into an incorrect one.

For the example program in \autoref{sec:example}, suppose the manually written tests \texttt{assertEquals(-1, f(0))}, \texttt{assertEquals(2, f(1))}, \texttt{assert Equals(8, f(7))}, and \texttt{assertEquals(7, f(8))} are provided. In this case, the synthesis process can already use the repair constraint imposed by these 4 tests to synthesize the correct condition if (x>0 \&\& x<8). Even if a bug-exposing test such as \texttt{assertEquals(4, f(5))} is generated, the repair constraint imposed by it will have a contradiction with the initial repair constraint (because it is impossible to synthesize a condition that satisfies the repair constraint imposed by all the 5 tests). Consequently, UnsatGuided will discard this bug-exposing test.

(2) \textbf{$P_{original}$ is A-overfitting}. In this case, $RC_{initial}$ is not strong enough to drive the synthesis-based repair techniques to synthesize a correct patch. More specifically, $RC_{initial}$ is in general strong enough to fully reflect the desired behaviors for correct input domain $I_{correct}$ but does not fully reflect the desired behaviors for all input points within buggy input domain $I_{bug}$. If the automatic test generation tool generates bug-exposing tests, the additional repair constraint enforced by a certain bug-exposing test does not necessarily have contradictions with $RC_{initial}$. If this happens, UnsatGuided is not able to identify and discard this kind of bug-exposing tests, and the synthesis process will be driven towards keeping the buggy behaviors corresponding to the bug-exposing tests. However, note this does not mean that the overfitting issue of incomplete fixing is worsened. If the behavior enforced by the kept bug-exposing test is already covered by the original patch, then it is likely that the synthesis process is not driven towards finding a new alternative solution and the overfitting issue of incomplete fixing remains the same. If the behavior enforced by the bug-exposing test is not covered by the original patch, then the synthesis process is likely to return a new solution. While the new solution indeed covers the new behavior enforced by the kept bug-exposing test, it can possibly generalize more over the whole $I_{bug}$ compared to the original patch. Thus, the overfitting issue of incomplete fixing can both be worsened and improved if a new solution is returned. Meanwhile, the normal tests generated by $T_{reg}$ by definition are not likely to be able to give additional repair constraints for input points within $I_{bug}$. 
Overall, for an A-overfitting patch, UnsatGuided is likely to have minimal positive impact and can coincidentally have a negative impact. 

To illustrate, assume the provided manually written tests are \texttt{assertEquals (-1, f(0))}, \texttt{assertEquals(2,f(1))}, \texttt{assertEquals(7,f(6))}, and \texttt{assert Equals(7, f(8))} for the example program in \autoref{sec:example}. Using the repair constraint enforced by these tests, the synthesis process can possibly synthesize the condition if (x>0 \&\& x<7), which is A-overfitting. Suppose bug-exposing test \texttt{assertE quals(4, f(5))} is generated, it will be discarded as the repair constraint imposed by it will make the synthesis process unable to synthesize a patch. However, if bug-exposing test \texttt{assertEquals(6, f(7))} is generated, it will be kept as there is no contradiction between the repair constraint enforced by it and that enforced by the manual tests and the synthesis process can successfully return a patch. In this specific case, even though the bug-exposing test is kept, the synthesized patch is not likely to change as the behavior enforced by the bug-exposing test is already covered by the original patch. In other words, the overfitting issue of incomplete fixing remains the same as the original patch.

(3) \textbf{$P_{original}$ is B-overfitting}. In this case, $RC_{initial}$ is also not strong enough to drive the synthesis-based repair techniques to synthesize a correct patch. In particular, $RC_{initial}$ is in general strong enough to fully reflect the desired behaviors for buggy input domain $I_{bug}$ but does not fully reflect the desired behaviors for all input points within correct input domain $I_{correct}$. In case the automatic test generation tool generates bug-exposing tests, $RC_{buggy}$ is likely to have contradictions with $RC_{initial}$ (note $RC_{initial}$ is in general strong enough for input points within $I_{bug}$). Thus, UnsatGuided will identify and discard these bug-exposing tests. Meanwhile, $RC_{normal}$ can supplement $RC_{initial}$ to better or even fully reflect the desired behaviors for input points within $I_{correct}$. Therefore, UnsatGuided can effectively help a B-overfitting patch reduce the overfitting issue of regression introduction, and can possibly turn a B-overfitting patch into a real correct one. 

For the example program in \autoref{sec:example}, assume the manually written tests \texttt{assertEquals(-1, f(0))}, \texttt{ assertEquals(2, f(1))}, \texttt{assertEquals(8, f(7))}, and \texttt{assertEquals(9, f(10))} are provided. Using the repair constraint enforced by these tests, the synthesis process can possibly synthesize the condition \texttt{if (x>0 \&\& x<10)}, which is B-overfitting. If bug-exposing test \texttt{assert Equals(5, f(6))} is generated, UnsatGuided will discard it as the repair constraint imposed by it will make the synthesis process unable to synthesize a patch. If a normal test such as \texttt{assertEquals(8, f(9))} is generated by $T_{reg}$, it provides additional repair constraint for input points within $I_{correct}$ and can possibly help the synthesis process to synthesize the condition \texttt{if (x>0 \&\& x<9)}, which has less overfitting issue of regression introduction compared to the original patch. In particular, if the normal test \texttt{assertEquals(7, f(8))} is generated by $T_{reg}$, this test will help the synthesis process to synthesize the exactly correct condition if (x>0 \&\& x<8).

(4) \textbf{$P_{original}$ is AB-overfitting}. This case is a combination of case (2) and case (3). UnsatGuided can effectively help an AB-overfitting patch reduce the overfitting issue of regression introduction, but has minimal positive impact on reducing the overfitting issue of incomplete fixing. Note as bug-exposing tests by definition are not likely to give additional repair constraints for input points within the correct input domain $I_{correct}$, so the strengthened repair constraints for input points within $I_{correct}$ are not likely to be impacted even if some bug-exposing tests are generated and not removed by UnsatGuided. In other words, UnsatGuided will still be effective in alleviating overfitting issue of regression introduction. 

Assume we have the manually written tests \texttt{assertEquals(-2, f(-1))}, \texttt{assertEquals(2, f(1))}, \texttt{assertEquals(7, f(6))}, and \texttt{assertEquals(7, f(8))} for the example program in \autoref{sec:example}. Using the repair constraint enforced by these tests, the synthesis process can possibly synthesis the condition if (x>-1 \&\& x<7), which is AB-overfitting. If bug-exposing test \texttt{assertEquals(6, f(7))} and normal test \texttt{assertEquals(-1, f(0))} are generated, both of them will be kept and the synthesis process can possibly synthesize the condition if (x>0 \&\& x<7), which has the same overfitting issue of incomplete fixing but less overfitting issue of regression introduction compared to the original patch.

In summary, UnsatGuided is not likely to break an already correct patch generated by a synthesis-based repair technique. For an overfitting patch, UnsatGuided can effectively reduce the overfitting issue of regression introduction, but has minimal positive impact on reducing the overfitting issue of incomplete fixing. With regard to turning an overfitting patch into a completely correct patch, UnsatGuided is likely to be effective only when the original patch generated using the manually written test suite is B-overfitting. 

\subsection{Discussion}
We now discuss the general usefulness of automatic test generation in alleviating overfitting for synthesis-based repair techniques. The overall conclusion is for techniques that make use of automatically generated tests to strengthen the repair constraint, there exists a fundamental limitation which makes the above core limitation of just effectively reducing the overfitting issue of regression introduction general, i.e., not specific to the proposed technique UnsatGuided. 

The fundamental limitation arises because of the oracle problem in automatic test generation. Due to the oracle problem, some of the automatically generated tests can encode wrong behaviors, which are called bug-exposing tests in this paper. Once the initial patch generated using the manually written test suite has the overfitting issue of incomplete fixing, the normal tests generated by an automatic test generation tool are not likely to be able to strengthen the repair constraints for input points within $I_{bug}$. While the bug-exposing tests generated by an automatic test generation tool can enforce additional repair constraints for input points within $I_{bug}$, the additional repair constraints enforced by bug-exposing tests are wrong. Different techniques can differ in how they classify automatically generated tests into normal tests and bug-exposing tests and how they further use these two kinds of tests, but they all face this fundamental problem. Consequently, for synthesis-based repair techniques, automatic test generation will not be very effective for alleviating the overfitting issue of incomplete fixing. 

However, for the overfitting issue of regression introduction, the normal tests generated by an automatic test case generation tool can effectively supplement the manually written test suite to better build the repair constraints for input points within $I_{correct}$. By using the strengthened repair constraint, synthesis-based repair techniques can synthesize a patch that has less or even no overfitting issue of regression introduction. \emph{According to this analysis, the usefulness of automatic test case generation in alleviating overfitting for synthesis-based repair techniques is mainly confined to reducing the overfitting issue of regression introduction.}

\section{Experimental Evaluation}
\label{sec:evaluation}

In this section, we present an empirical evaluation of the effectiveness of UnsatGuided in alleviating overfitting problems for synthesis-based repair techniques. 
In particular, we aim to empirically answer the following research questions:

\begin{itemize}
\item \textbf{RQ1}: How frequently do overfitting issues of incomplete fixing and regression introduction occur in practice for synthesis-based repair techniques?

\item \textbf{RQ2}: How does UnsatGuided perform with respect to alleviating overfitting issues of incomplete fixing and regression introduction? 

\item \textbf{RQ3}: What is the impact of UnsatGuided on the correctness of the patches?

\item \textbf{RQ4}: How does UnsatGuided respond to bug-exposing tests?

\item \textbf{RQ5}: What is the time overhead of UnsatGuided?

\end{itemize}

\subsection{Subjects of Investigation}

\subsubsection{Subject Programs}

We selected Defects4J (\cite{JustJE2014}), a known database of real faults from real-world Java programs, as the experimental benchmark. Defects4J has different versions and the latest version of the benchmark contains 395 faults from 6 open source projects. Each fault in Defects4J is accompanied by a manually written test suite which contains at least one test that exposes the fault. In addition, Defects4J also provides commands to easily access faulty and fixed program versions for each fault, making it relatively easy to analyze them. Among the 6 projects, Mockito has been configured and added to the Defects4J framework recently (after we start the study presented in this paper). Thus we do not include the 38 faults for Mockito in our study. Besides, we also discard the 133 faults for Closure compiler as the tests are organized using scripts rather than the standard JUnit tests, which prevents these tests from running within our repair infrastructure. Consequently, we use the 224 faults of the remaining 4 projects in our experimental evaluation. \autoref{tab:dataset} gives basic information about these 4 subjects.

\begin{table}
 \caption{Descriptive Statistics of the 224 Considered Faults in Defects4J}
  \label{tab:dataset}
  \small
  \begin{tabular}{|l|r|r|r|r|r|}
    \hline
    Subjects & \#Bugs & \tabincell{c}{Source\\KLoC} & \tabincell{c}{Test\\KLoC} & \#Tests & Dev years \\
    \hline
    JFreechart   & 26  & 96 & 50 & 2,205 & 10 \\
    Commons Math & 106 & 85 & 19 & 3,602 & 14 \\
    Joda-Time    & 27  & 28 & 53 & 4,130 & 14 \\
    Common Lang  & 65  & 22 &  6 & 2,245 & 15 \\
    \hline
\end{tabular}
\end{table}

\subsubsection{Synthesis-based Repair Techniques}

For our approach UnsatGuided to be implemented, we need a stable synthesis-based repair technique. In this study, Nopol (\cite{nopol}) is used as the representative of synthesis-based repair techniques. We select it for two reasons. First, Nopol is the only publicly-available synthesis-based repair technique that targets modern Java code. Second, it has been shown that Nopol is an effective automated repair system that can tackle real-life faults in real-world programs (\cite{defects4j-repair}).
  
\subsubsection{Automatic Test Case Generation Tool}
\label{sec:ATCG}

The automatic test case generation tool used in this study is EvoSuite (\cite{ESECFSE11}). EvoSuite aims to generate tests with maximal code coverage by applying a genetic algorithm. Starting with a set of random tests, it then uses a coverage based fitness function to iteratively apply typical search operators such as selection, mutation, and crossover to evolve them. Upon finishing the search, it minimizes the test suite with highest code coverage with respect to the coverage criterion and adds regression test assertions. 
To our knowledge, EvoSuite is the state-of-art open source Java unit test generation tool. Compared with another popular test generation tool Randoop (\cite{randoop}), some recent studies (\cite{evidence, 7372009}) have shown that Evosuite is better than Randoop in terms of a) compilable test generated, b) minimized flakiness, c) false positives, d) coverage, and e) most importantly--the number of bugs detected.
While the generated tests by EvoSuite can possibly have problems of creating complex objects, exposing complex conditions, accessing private methods or fields, creating complex interactions, and generating appropriate assertions, they can be considered as effective in finding bugs in open-source and industrial systems in general (\cite{7372009}). 
Besides, as shown in algorithm 1, the approach UnsatGuided requires that the automatic test case generation tool is able to target a specific file of the program under repair. EvoSuite is indeed capable of generating tests for a specific class. 

To generate more tests and make the test generation process itself as deterministic as possible, i.e., the generated tests should be the same if somebody else repeats out experiment, we made some changes about the timeout value, search budget value, sandboxing and mocking setting in the default EvoSuite option. The complete EvoSuite setting is available on Github.\footnote{\url{https://github.com/Spirals-Team/test4repair-experiments}} 

\subsection{Experimental Setup}
\label{sec:setup}

For each of the 224 studied faults in the Defects4J dataset, we run the proposed approach UnsatGuided against it. Whenever the test generation process is invoked, we run EvoSuite 30 times with different seeds to account for the randomness of EvoSuite following the guideline given in (\cite{Arcuri2011}). The 30 seeds are 30 integer numbers randomly selected between 1 and 200. In addition, EvoSuite can generate tests that do not compile or generates tests that are unstable (i.e., tests which could fail or pass for the same configuration) due to the use of non-deterministic APIs such as date and time of day. Similar to the work in (\cite{JustJIEHF2014,7372009}), we use the following process to remove the uncompilable and unstable tests if they exist:
\begin{enumerate}[label=(\roman*)]
\item Remove all uncompilable tests;
\vspace{1mm} 
\item Remove all tests that fail during re-execution on the program to be repaired;
\vspace{1mm} 
\item Iteratively remove all unstable tests: we execute each compliable test suite on the program to be repaired five times consecutively. If any of these executions reveals unstable tests, we then remove these tests and re-compile and re-execute the test suite. This process is repeated until all remaining tests in the test suite pass five times consecutively. 
\end{enumerate}

Our experiment is extremely time-consuming. To make the time cost manageable, the timeout value for UnsatGuided, i.e., the input time budget in algorithm 1 for Nopol, is set to be 40 minutes in our experimental evaluation. Besides this change to global timeout value, we use the default configuration parameters of Nopol during its run. The experiment was run on a cluster consisting of 200 virtual nodes running Ubuntu 16.04 on a single Intel 2.68 GHz Xeon core with 1GB of RAM. As UnsatGuided will invoke the synthesis-based repair technique for each test generated, the whole repair process may still cost a lot of time. If so, we reduce the number of considered seeds. This happens for 2 faults (Chart\_26 and Math\_24), for which combining Nopol with UnsatGuided will generally cost more than 13 hours for each EvoSuite seed. Consequently, we use 10 seeds for these two bugs only for sake of time.
Following an open-science ethics, all the code and data is made publicly available on the mentioned Github site in \autoref{sec:ATCG}.

\subsection{Evaluation Protocol} 
\label{sec:protocol}

We evaluate the effectiveness of UnsatGuided from two points: its impact on the overfitting issue and correctness of the original patch generated by Nopol.

\subsubsection{Assess Impact on Overfitting Issue}
\label{sec:overfittingissue}
We have several major phases to evaluate the impact of UnsatGuided on overfitting issue of the original Nopol patch.

(1) \emph{Test Case Selection and Classification}. To determine whether a patch has overfitting issue of incomplete fixing or regression introduction, we need to see whether the corresponding patched program will fail tests from buggy input domain $I_{bug}$ or correct input domain $I_{correct}$ of the program to be repaired. As it is impractical to enumerate all tests from these two input domains, we view all tests generated for all seeds during our run of UnsatGuided (see \autoref{sec:setup}) for a buggy program version as a representative subset of tests from these two input domains for this buggy program version in this paper. We believe it is reasonable from two aspects. On the one hand, we use a large number of seeds (30 in most cases) for each buggy program version, so we will have a large number of tests in general for each buggy program version. On the other hand, these tests all focus on testing the behaviors related with the patched highly suspicious files.

We then need to classify the generated tests as being in the buggy input domain or being in the correct input domain. Recall that during our run of UnsatGuided, EvoSuite uses the version-to-be-repaired as the oracle to generate tests. After the run of UnsatGuided for each seed, we thus have an EvoSuite test set which contain both 1) normal tests whose inputs are from $I_{correct}$ and the assertions of them are right, and 2) bug-exposing tests whose inputs are from $I_{bug}$ and the assertions of them are wrong. To distinguish these two kinds of tests, we use the correct version of the version-to-be-repaired to achieve this goal. Note the assumption of the existence of a correct version is used here just for the evaluation purpose, we do not have this assumption for the run of UnsatGuided.

More specifically, given a buggy program $P_{buggy}$, the correct version $P_{correct}$ of $P_{buggy}$, and an EvoSuite test suite $TS_{Evo\_i}$ generated during the run of UnsatGuided for seed $seed_{i}$, 
we run $TS_{Evo\_i}$ against $P_{correct}$ to identify bug-exposing tests. As $TS_{Evo\_i}$ is generated from $P_{buggy}$, tests can possibly assert wrong behaviors. Thus, \emph{a test fails over $P_{correct}$ is a bug-exposing test} and is added to the test set $TS_{bugexpo}$. Otherwise, it is a normal test and is added to the test set $TS_{normal}$. For a certain buggy program version, this process is executed for each EvoSuite test suite $TS_{Evo\_j}$ generated for each seed $seed_{j}$ of the seed set $\{seed_{j}|1\leqslant j\leqslant N, N=30\; or \;10\}$. Consequently, for a specific buggy program version, $TS_{bugexpo}$ contains all bug-exposing tests and $TS_{normal}$ contains all normal tests among all tests generated for all seeds during the run of UnsatGuided for this buggy program version. 

(2) \emph{Analyze the Overfitting Issue of the Synthesized Patches}. For a buggy program $P_{buggy}$, the correct version $P_{correct}$ of $P_{buggy}$, and the patch $pc$ to $P_{buggy}$, we then use the identified test sets $TS_{bugexpo}$ and $TS_{normal}$ in the previous step to analyze the overfitting issue of $pc$.

To determine whether patch $pc$ has overfitting issue of regression introduction, we execute the program obtained by patching buggy program $P_{buggy}$ with $pc$ against $TS_{normal}$. If at least one test in $TS_{normal}$ fails, then patch $pc$ has overfitting issue of regression introduction.

To determine whether patch $pc$ has overfitting issue of incomplete fixing, it is harder. The basic idea is executing the program obtained by patching buggy program $P_{buggy}$ with $pc$ against $TS_{bugexpo}$, and patch $pc$ has overfitting issue of incomplete fixing if at least one test in $TS_{bugexpo}$ fails. However, recall that the tests in $TS_{bugexpo}$ are generated based on the buggy version $P_{buggy}$, i.e., the oracles are incorrect. Consequently, we first need to obtain the correct oracles for all tests in $TS_{bugexpo}$. We again use the correct version $P_{correct}$ to achieve this goal and the process is as follows. 

First, for each failing assertion contained in a test from $TS_{bugexpo}$, we first capture the value it receives when the test is executed on the correct version $P_{correct}$. 
For instance, given a failing assertion \texttt{assertEquals(10,calculateValue($y$))}, 10 is the value that the assertion expects and the value from \texttt{calculateValue($y$)} is the received value. 
For this specific example, we need to capture the value for \texttt{calculateValue($y$)} on $P_{correct}$ (note the value that $P_{buggy}$ returns for \texttt{calculateValue($y$)} is 10). 
Then, we replace the expected value in the failing assertion with the received value established on $P_{correct}$. 
For the previous example, if \texttt{calculateValue($y$)} returns the value 5 on $P_{correct}$, the repaired assertion is \texttt{assertEquals(5, calculateValue($y$))}.

The above process turns $TS_{bugexpo}$ into $TS_{bugexpo{\cmark}}$ so that all bug-exposing tests will have correct oracles. After obtaining $TS_{bugexpo{\cmark}}$, we run $TS_{bugexpo{\cmark}}$ against the program obtained by patching buggy program $P_{buggy}$ with $pc$. If we observe any failing tests, then patch $pc$ has overfitting issue of incomplete fixing. 

(3) \emph{Measure Impact}. To evaluate the impact of UnsatGuided on the overfitting issue for a certain buggy program version, we compare the overfitting issue of the original Nopol patch $pc_{original}$ generated using the manually written test suite with that of the new patch $pc_{new}$ generated after running UnsatGuided. More specifically, the process is as follows. 

First, we use phases (1) and (2) to see whether the original patch $pc_{original}$ has overfitting issue of incomplete fixing or regression introduction. When we observe failing tests from $TS_{normal}$ or $TS_{bugexpo{\cmark}}$, we record the detailed number of failing tests. The recorded number represents the severity of the overfitting issue.  

Second, for a patch $pc_{new\_i}$ generated by running UnsatGuided using a certain seed $seed_{i}$, we also use phases (1) and (2) to see whether the new patch $pc_{new\_i}$ has overfitting issue of incomplete fixing or regression introduction and record the number of failing tests if we observe failing tests from $TS_{normal}$ or $TS_{bugexpo{\cmark}}$. Note besides the test suite (corresponding to $seed_{i}$) used by UnsatGuided to generate $pc_{new\_i}$, we also use all the other test suites generated for other seeds to evaluate the overfitting issue of $pc_{new\_i}$. 

Finally, the result obtained for $pc_{new\_i}$ is compared with that for $pc_{original}$ to determine the impact of UnsatGuided. 

We repeat this process for each patch generated using each seed for a certain program version (i.e., the patch set \{$pc_{new\_i}$ $|1\leqslant i\leqslant N, N=30\;or\;10\}$), and use the average result to assess the overall impact of UnsatGuided. 

\subsubsection{Assess Impact on Correctness}
We compare the correctness of the patch generated after the run of UnsatGuided with that generated using Nopol to see the impact of UnsatGuided on patch correctness. To determine the correctness of a patch, the process is as follows.

First, we look at whether the generated tests reveal that there exist overfitting issues for a certain generated patch according to the procedure in \autoref{sec:overfittingissue}.
 
Second, we manually analyze the generated patch and compare it with the corresponding human patch. A generated patch is deemed as correct only if it is exactly the same or semantically equivalent to the human patch. The equivalence is established based on the authors' understanding of the patch. To reduce the possible bias introduced as much as possible, two of the authors analyze the correctness of the patches separately and the results reported in this paper are based on the agreement between them. Note that the corresponding developer patches for several buggy versions trigger exceptions and emit text error messages if certain conditions are true, we count a generated patch correct if it triggers the same type of exceptions as the human patch under the exception conditions and we do not take the error message into account.

Note due to the use of different Nopol versions, the Nopol patches generated in this paper for some buggy versions are different from that generated in (\cite{defects4j-repair}). We thus replicate the manual analysis of the original Nopol patches. 

As we use a large number of seeds (30 in most cases) for running UnsatGuided, it can happen that we have a large number of generated patches that are different from the original Nopol patch for a certain buggy version. For the inherent difficulty of the manual analysis, it is unrealistic to analyze all of the newly generated patches. To make the manual analysis realistic, for each buggy version, we randomly select one patch that is different from the original Nopol patch across all of the different kinds of patches generated for all seeds. It can happen that for a certain buggy version, the newly generated patches after the run of UnsatGuided for all seeds are the same as the original Nopol patch. In this case, it is obvious that UnsatGuided has no impact on the change of patch correctness. 

\subsection{Result Presentation}
\autoref{tab:nopol-results} displays the experimental results on combining Nopol with UnsatGuided (hereafter referred to as Nopol+\-Unsat\-Guided). This table only shows the Defects4J bugs that can be originally repaired by Nopol, and their identifiers are listed in column \emph{Bug ID}. 

\begin{landscape}
\begin{table*}[htbp]
  \tiny
  \caption{Experimental results with Nopol+UnsatGuided on the Defects4j Repository, only show bugs with test-suite adequate patches by plain Nopol.}
  \centering
  \label{tab:nopol-results}
  \begin{tabular}{|l|c|c|c|c|c|r|r|r|r|r|c|r|r|r|r|}
    \hline
    \multirow{7}{*}{\rotatebox{-90}{Bug ID}} & \multicolumn{2}{c|}{Tests} & \multicolumn{4}{c|}{Nopol}  & \multicolumn{7}{c|}{Nopol+UnsatGuided} \\ 
    \cline{2-14}
     & \rotatebox{-90}{\#EvoTests} & \rotatebox{-90}{\#Bug-expo} & \rotatebox{-90}{Time (hh:mm)} & \rotatebox{-90}{\tabincell{c}{incomplete fix \\ (\#failing)}} & \rotatebox{-90}{regression (\#failing)} & \rotatebox{-90}{correctness} & \rotatebox{-90}{\#Removed} & \rotatebox{-90}{\#Removed Bug-expo} & \rotatebox{-90}{Avg \#Time (hh:mm)} & \rotatebox{-90}{\tabincell{c}{Change ratio \\ (\#unique)}} & \rotatebox{-90}{\tabincell{c}{fix completeness \\ change \\ (Avg \#Removedinc)}} & \rotatebox{-90}{\tabincell{c}{regression change\\ (Avg \#Removedreg)}} & \rotatebox{-90}{correctness} \\
    \hline
Chart\_1 & 3012 & 0 & 00:02 & No (0) & No (0) & NO & 0 & 0 & 03:00 & 0/30 (1) &  same (0) & same (0) & NO\\
Chart\_5 & 2931 & 3 & 00:01 & No (0) & Yes (10) & NO & 104 & 3 & 01:18 & 27/30 (27) & same (0) & improve (2.9) & NO\\
Chart\_9 & 3165 & 0 & 00:01 & No (0) & No (0) & NO & 0 & 0 & 01:00 & 0/30 (1) & same (0) & same (0) & NO\\
Chart\_13 & 852 & 0 & 00:02 & No (0) & No (0) & NO & 0 & 0 & 00:24 & 30/30 (2) &  same (0) & same (0) & NO\\
Chart\_15 & 3711 & 0 & 00:04 & No (0) & Yes (4) & NO & 5 & 0 & 06:48 & 27/30 (23) & same (0) & improve (2.0) & NO\\
Chart\_17 & 3246 & 10 & 00:01 & Yes (10) & No (0) & NO & 27 & 0 & 00:48 & 0/30 (1)  & same (0) & same (0) & NO\\
Chart\_21 & 1584  & 0 & 00:01 & No (0) & Yes (6) & NO & 0 & 0 & 00:48 & 30/30 (30) & same (0) & improve (6.0)$\star$ & NO\\
Chart\_25 & 441 & 0 & 00:01 & No (0) & Yes (8) & NO & 0 & 0 & 00:12 & 8/30 (6) &  same (0) & improve (8.0)$\star$ & NO\\
Chart\_26 & 2432 & 0 & 00:03 & No (0) & Yes (6) & NO & 6 & 0 & 13:36 & 10/10 (5) &  same (0) & improve (6.0)$\star$ & NO \\
Lang\_44 & 3039 & 13 & 00:01 & No (0) & No (0) & \bf{YES} & 13 & 13 & 00:48 & 3/30 (2) & same (0) & same (0) & \bf{YES}\\
Lang\_51 & 3720 & 1 & 00:01 & No (0) & No (0) & NO & 15 & 1 & 01:00 & 29/30 (2) &  same (0) & same (0) & NO\\
Lang\_53 & 2931 & 0 & 00:01 & No (0) & No (0) & NO & 0 & 0 & 00:06 & 26/30 (18) &  same (0) &  same (0) & NO\\
Lang\_55 & 606 & 0 & 00:01 & No (0) & No (0) & \bf{YES} & 1 & 0 & 00:12 & 30/30 (1) & same (0) & same (0) & \bf{YES}\\
Lang\_58 & 6471 & 0 & 00:01 & No (0) & Yes (5) & NO & 5 & 0 & 01:42 & 0/30 (1) &  same (0) & same (0) & NO\\
Lang\_63 & 1383 & 1 & 00:01 & No (0) & No (0) & NO & 33 & 1 & 00:36 & 27/30 (5) & same (0) & same (0) & NO\\
Math\_7 & 876 & 2 & 00:16 & Yes (2) & No (0) & NO & 0 & 0 & 05:00 & 2/30 (3) & same (0) &  same (0) & NO\\
Math\_24 & 1327 & 0 & 00:15 & No (0) & No (0) & NO & 25 & 0 & 24:06 & 10/10 (10) & same (0) & same (0) & NO\\
Math\_28 & 219 & 0 & 00:17 & No (0) & No (0) & NO & 0 & 0 & 00:30 & 0/30 (1) & same (0) & same (0) & NO\\
Math\_33 & 1749 & 1 & 00:13 & Yes (1) & No (0) & NO & 19 & 0 & 10:30 & 28/30 (8) & same (0) & worse (-2.0) & NO\\
Math\_40 & 831 & 71 & 00:16 & Yes (71) & Yes (21) & NO & 392 & 0 & 07:00 & 7/30 (8) & same (0) & same (0) & NO\\
Math\_41 & 1224 & 0 & 00:06 & No (0) & Yes (41) & NO  & 35 & 0 & 02:00 & 27/30 (27) &  same (0) & improve (35.1) & NO\\
Math\_42 & 1770 & 19 & 00:04 & Yes (19) & No (0) & NO & 2 & 0 & 03:54 & 24/30 (22) &  same (0) & same (0) & NO\\
Math\_50 & 1107 & 26 & 00:11 & Yes (21) & Yes (45) & NO  & 23 & 1 & 04:36 & 28/30 (27) & improve (1.1) & improve (41.0) & NO\\
Math\_57 & 651 & 0 & 00:03 & No (0) & No (0) & NO & 0 & 0 & 00:48 & 15/30 (4) &  same (0) & same (0) & NO\\
Math\_58 & 228 & 0 & 00:06 & No (0) & No (0) & NO & 7 & 0 & 00:20 & 2/30 (2) & same (0) & same (0) & NO\\
Math\_69 & 897 & 0 & 00:01 & No (0) & No (0) & NO & 30 & 0 & 00:12 & 30/30 (21) & same (0) & same (0) & NO\\
Math\_71 & 951 & 0 & 00:01 & No (0) & Yes (56) & NO & 17 & 0 & 00:24 & 25/30 (11) & same (0) & improve (53.0) & NO\\
Math\_73 & 1035 & 0 & 00:01 & No (0) & Yes (1) & NO & 10 & 0 & 00:18 & 25/30 (24) & same (0) & improve (1)$\star$ & NO\\
Math\_78 & 1014 & 0 & 00:01 & No (0) & Yes (44) & NO & 49 & 0 & 00:24 & 28/30 (16) & same (0) & improve (34.9) & NO\\
Math\_80 & 1356 & 67 & 00:01 & Yes (49) & No (0) & NO & 29 & 1 & 00:54 & 29/30 (27) & worse (-17.9) & same (0) & NO\\
Math\_81 & 1320 & 4 & 00:01 & Yes (4) & Yes (35) & NO & 30 & 0 & 00:24 & 23/30 (22) & same (0) & improve (35.0)$\star$ & NO\\
Math\_82 & 510 & 0 & 00:01 & No (0) & No (0) & NO & 0 & 0 & 00:08 & 0/30 (1) & same (0) & same (0) & NO\\
Math\_84 & 165 & 0 & 00:01 & No (0) & No (0) & NO & 0 & 0 & 00:06 & 0/30 (1) & same (0) & same (0) & NO\\
Math\_85 & 798 & 0 & 00:01 & No (0) & No (0) & NO & 32 & 0 & 00:12 & 28/30 (11) & same (0) & same (0) & \bf{YES}\\
Math\_87 & 1866 & 14 & 00:01 & Yes (13) & Yes (8) & NO & 0 & 0 & 00:54 & 29/30 (29) & worse (-1) & improve (8.0)$\star$ & NO\\
Math\_88 & 1890 & 11 & 00:01 & Yes (11) & No (0) & NO & 0 & 0 & 00:30 & 06/30 (7) & same (0) & same (0) & NO\\
Math\_105 & 1353 & 7 & 00:09 & Yes (7) & Yes (6) & NO & 6 & 0 & 04:20 & 29/30 (30) & same (0) & improve (2.9) & NO\\
Time\_4 & 2778 & 5 & 00:01 & Yes (5) & Yes (6) & NO & 0 & 0 & 00:54 & 23/30 (23) & improve (0.8) & improve (5.7) & NO\\
Time\_7 & 1491 & 0 & 00:01 & No (0) & Yes (11) & NO & 12 & 0 & 00:54 & 12/30 (13) & same (0) & worse (-1) & NO\\
Time\_11 & 1497 & 5 & 00:04 & Yes (5) & No (0) & NO & 7 & 0 & 01:36 & 0/30 (1) & same (0) & same (0) & NO\\
Time\_14 & 687 & 0 & 00:01 & No (0) & Yes (3) & NO & 1 & 0 & 00:18 & 24/30 (23) & same (0) & improve (2.0) & NO\\
Time\_16 & 1476 & 0 & 00:01 & No (0) & Yes (6) & NO & 5 & 0 & 00:24 & 1/30 (2) & same (0) & improve (1) & NO\\
    \hline
  \end{tabular}
\end{table*}

\end{landscape}

The test generation results by running EvoSuite are shown in the two columns under the column \emph{Tests}, among which the \emph{\#EvoTests} column shows the total number of tests generated by EvoSuite for all seeds and the \emph{\#Bug-expo} column shows the number of bug-exposing tests among all of the generated tests. 

The results obtained by running just Nopol are shown in the columns under the column \emph{Nopol}. The \emph{Time} column shows the time used by Nopol to generate the initial patch. The \emph{incomplete fix (\#failing)} column shows what is the overfitting issue of incomplete fixing for the original Nopol patch. Each cell in this column is of the form X (Y), where X can be ``Yes'' or ``No'' and Y is a digit number. The ``Yes'' and ``No'' mean that the original Nopol patch has and does not have overfitting issue of incomplete fixing respectively. The digit number in parentheses shows the number of bug-exposing tests on which the original Nopol patch fails. Similarly, the \emph{regression (\#failing)} column tells what is the overfitting issue of regression introduction for the original Nopol patch, and each cell in this column is of the same form with the column \emph{incomplete fix (\#failing)}. The ``Yes'' and ``No'' for this column mean that the original Nopol patch has and does not have overfitting issue of regression introduction respectively. The digit number in parentheses shows the number of normal tests on which the original Nopol patch fails. Finally, the column \emph{correctness} shows whether the original Nopol patch is correct, with ``Yes'' representing correct and ``No'' representing incorrect.

The results obtained by running Nopol+UnsatGuided are shown in the remaining columns under the column \emph{Nopol}+\emph{UnsatGuided}. 
The \emph{\#Removed} column shows the total number of removed generated tests during the run of Nopol+UnsatGuided for all seeds. 
The number of bug-exposing tests among the removed tests is shown in the column \emph{\#Removed Bug-expo}. 
The \emph{Avg\#Time} column shows the average time used by Nopol+UnsatGuided to generate the patch for each seed. The \emph{Change ratio (\#unique)} column is of the form \emph{X}/\emph{Y} (\emph{Z}). 
Here \emph{Y} is the number of different seeds used, \emph{X} refers to the number of generated patches by Nopol+UnsatGuided that are different from the original Nopol patch, and \emph{Z} is the number of distinct patches among all of the patches generated for all seeds. 

The following two columns \emph{fix completeness change (Avg\#Removedinc)} and \emph{regression change (Avg\#Removedreg)} show the effectiveness of UnsatGuided in alleviating overfitting issue of incomplete fixing and regression introduction respectively. 
Each cell in these two columns is of the form X (Y), where X can be ``improve'', ``worse'', and ``same'' and Y is a digit number. 

Compared with the original Nopol patch, the ``improve'', ``worse'', and ``same'' in column \emph{fix completeness change (Avg\#Removedinc)} mean that the new patch generated by running Nopol+\-Unsat\-Guided has less, more, and the same overfitting issue of incomplete fixing respectively. The digit number gives a more detailed information. In particular, it gives the average number of removed failing bug-exposing tests for the new patch generated by running Nopol+UnsatGuided compared with the original Nopol patch. In other words, the digital value is obtained by subtracting the average number of failing bug-exposing tests for the new patch generated by running Nopol+UnsatGuided from the number of failing bug-exposing tests for the original Nopol patch.
A positive value is good, which shows that the new patch has less overfitting issue of incomplete fixing in a way. For example, a value of 1 says that the new patch does not exhibit overfitting issue of incomplete fixing anymore for a test case within $I_{bug}$.   

Similarly, compared with the original Nopol patch, the ``improve'', ``worse'', and ``same'' in column \emph{regression change (Avg\#Removedreg)} mean that the new patch generated by running Nopol+\-Unsat\-Guided has less, more, and the same overfitting issue of regression introduction respectively. Compared with the original Nopol patch, the digit number in column \emph{regression change (Avg\#Removedreg)} gives the average number of removed failing normal tests for the new patch generated by running Nopol+UnsatGuided, and it equates to the value obtained by subtracting the average number of failing normal tests for the new patch generated by running Nopol+UnsatGuided from the number of failing normal tests for the original Nopol patch. Again, a positive value is good, which shows that the new patch has less overfitting issue of regression introduction in a way. For example, a value of 2 says that the new patch does not exhibit overfitting issue of regression introduction anymore for two test cases within $I_{correct}$. 

Note for the patch generated using Nopol+UnsatGuided for a certain seed, the tests considered are all tests generated using all seeds for the corresponding program version. We average the results for all seeds of a certain program version and the resultant numbers are shown as digit numbers in the columns \emph{fix completeness change (Avg\#Removedinc)} and \emph{regression change (Avg\#Removedreg)}. Overall, \emph{a positive digit number in these two columns shows an improvement: it means that overfitting issue of incomplete fixing or regression introduction has been alleviated after running UnsatGuided}. In addition, we use ``perfect'' to refer to the situation where for each seed of a certain program version, running \emph{Nopol}+\emph{UnsatGuided} with the seed will get a patch that will completely remove the overfitting issue of the original Nopol patch. The ``perfect'' results are illustrated with $(\star)$.

Finally, the column \emph{correctness} under the column \emph{Nopol}+\emph{UnsatGuided} shows whether the selected patch generated by running \emph{Nopol}+\emph{UnsatGuided} is correct, again with ``Yes'' representing correct and ``No'' representing incorrect.

\subsection{\textbf{RQ1}: Prevalence of the Two Kinds of Overfitting Issues}
We first want to measure the prevalence of overfitting issues of incomplete fixing and regression introduction among the patches generated by synthesis-based repair techniques.

We can see from the \emph{incomplete fix (\#failing)} and \emph{regression (\#failing)} columns under the column \emph{Nopol} that for the 42 buggy versions that Nopol can generate an initial patch, overfitting can be observed for 26 buggy versions (when there exists ``Yes'' in either of these two columns). 

Among the other 16 buggy versions for which we do not observe any kinds of overfitting issues, the manual analysis shows that the Nopol patches for two buggy versions (Lang\_44 and Lang\_55) are correct. However, the manual analysis shows that the Nopol patches for the remaining 14 buggy versions are incorrect, yet we do not observe any number of failing bug-exposing or normal tests for the programs patched with the patches generated by Nopol. This shows the limitation of automatic test case generation in covering the buggy input domain $I_{bug}$ for real programs, which confirms a previous study (\cite{7372009}).

Among the 26 buggy versions for which we observe overfitting issues, the original Nopol patches for 13 buggy versions have the overfitting issue of incomplete fixing, the original Nopol patches for 19 buggy versions have the overfitting issue of regression introduction, and the original Nopol patches for 6 buggy versions have both the overfitting issues of incomplete fixing and regression introduction. Thus, both the overfitting issues of incomplete fixing and regression introduction are common for the Nopol patches. 

It can also be seen from \autoref{tab:nopol-results} that the severity of overfitting differs from one patch to another as measured by the number of failing tests. Among the 13 patches that have overfitting issue of incomplete fixing, the number of failing bug-exposing tests is less than 3 for 3 patches (which implies the overfitting issue is relatively light), yet this number is larger than 20 for 3 patches (which implies the overfitting issue is relatively serious). 

Similarly, for the 19 patches that have overfitting issue of regression introduction, the number of failing normal tests is less than 3 for 1 patch (which implies the overfitting issue is relatively light), yet this number is larger than 20 for 6 patches (which implies the overfitting issue is relatively serious). 

\begin{mdframed}
\textbf{Answer for RQ1}: Both overfitting issues of incomplete fixing (13 patches) and regression introduction (19 patches) are common for the patches generated by Nopol. 
\end{mdframed}

\subsection{\textbf{RQ2}: Effectiveness of UnsatGuided in Alleviating Overfitting Issues}
We then want to assess the effectiveness of UnsatGuided. It can be seen from the column \emph{Change ratio (\#unique)} of \autoref{tab:nopol-results} that for the 42 buggy versions that can be initially repaired by Nopol, the patches generated for 34 buggy versions have been changed at least for one seed after running Nopol+UnsatGuided. If we consider all executions (one per seed per buggy version), we obtain a total of 1220 patches with Nopol+UnsatGuided. Among the 1220 patches, 702 patches are different from the original patches generated by running Nopol only. Thus, UnsatGuided can significantly impact the output of the Nopol repair process. We will further investigate the quality difference between the new Nopol+UnsatGuided patches and the original Nopol patches. 

The results for alleviating the two kinds of overfitting issues by running Nopol+ UnsatGuided are displayed in the columns \emph{fix completeness change (Avg \#Removedinc)} and \emph{regression change (Avg\#Removedreg)} of \autoref{tab:nopol-results}. 

With regard to alleviating the overfitting issue of incomplete fixing, we can see from the column \emph{fix completeness change (Avg\#Removedinc)} that UnsatGuided has an effect on 4 buggy program versions (Math\_50, Math\_80, Math\_87 and Time\_4). For all those 4 buggy versions, the original Nopol patch already has the overfitting issue of incomplete fixing. With UnsatGuided, the overfitting issue of incomplete fixing has been alleviated in 2 cases (Math\_50, Time\_4) and worsened for 2 other cases (Math\_80, Math\_87). This means UnsatGuided is likely to have a minimal positive impact on alleviating overfitting issue of incomplete fixing and can possibly have a negative impact on it, confirming our analysis in \autoref{sec:analysis}. 
We will further discuss this point in RQ4 (\autoref{sec:answer-rq4}). 

In terms of alleviating overfitting issue of regression introduction, we can see from the column \emph{regression change (Avg\#Removedreg)} that UnsatGuided has an effect on 18 buggy program versions. Among the 18 original Nopol patches for these 18 buggy program versions, UnsatGuided has alleviated the overfitting issue of regression introduction for 16 patches. In addition, for 6 buggy program versions, the overfitting issue of regression introduction of the original Nopol patch has been completely removed.  These 6 cases are indicated with $(\star)$ in \autoref{tab:nopol-results}. Meanwhile, UnsatGuided worsens the overfitting issue of regression introduction for two other original Nopol patches (Math\_33 and Time\_7). 
It can possibly happen as even though the repair constraint for input points within $I_{correct}$ has been somewhat strengthened (but not completely correct), yet the solution of the constraint happens to be more convoluted.
Overall, with 16 positive versus 2 negative cases, UnsatGuided can be considered as effective in alleviating overfitting issue of regression introduction.

\begin{mdframed}
\textbf{Answer for RQ2}:  UnsatGuided can effectively alleviate the overfitting issue of regression introduction (16/19 cases), but has minimal positive impact on reducing the overfitting issue of incomplete fixing.
This results confirm our deductive analysis of the effectiveness of UnsatGuided in alleviating the two kinds of overfitting issues (\autoref{sec:analysis}).
\end{mdframed}

\subsection{\textbf{RQ3}: Impact of UnsatGuided on Patch Correctness} 
We will further assess the impact of UnsatGuided on the correctness of the patches. More specifically, we will assess 1) whether running Nopol+UnsatGuided destroys the already correct patches generated by Nopol (i.e., make them become incorrect) and 2) whether running Nopol+UnsatGuided can change an overfitting patch generated by Nopol into a completely correct one. 
    
\textbf{Can already correct patches be broken?}
The previous paper (\cite{defects4j-repair}) claims that running Nopol can generate correct patches for 5 buggy program versions Chart\_5, Lang\_44, Lang\_55, Lang\_58, and Math\_50. However, for three of them (Chart\_5, Lang\_58, and Math\_50), we can see from \autoref{tab:nopol-results} that some EvoSuite tests fail on the original Nopol patches. Due to the use of different Nopol versions, the Nopol patch generated in this paper for Math\_50 is different from that in (\cite{defects4j-repair}). We run the EvoSuite tests against the Nopol patch in (\cite{defects4j-repair}) and we also observe failing tests. To ensure the validity of the bug detection results, two authors of this paper have manually checked the correctness of the patches generated for these three buggy versions in the paper (\cite{defects4j-repair}). The overall results suggest that the original Nopol patches for these three program versions are not truly correct, which shows the inherent difficulty of manual analysis. For the other 2 buggy program versions (Lang\_44 and Lang\_55), there is no indication of overfitting and we consider the original Nopol patches as well as the new patches generated by running Nopol+UnsatGuided as correct. 
We now demonstrate why they can be considered as correct.

For Lang\_44, the bug arises for a method which parses a string to a number (\texttt{String} to \texttt{int}, \texttt{long}, \texttt{float} or \texttt{double}) (see \autoref{fig:case1}). If the string (\emph{val}) only contains the char \emph{L} which specifies the type \texttt{long}, the method returns an \emph{IndexOutOfBoundsException} (due to the expression \texttt{numeric.substring(1)} in the \emph{if} condition) instead of the expected \emph{NumberFormatException}, the other situations have already been correctly handled. The human patch adds a check at the beginning of the method to avoid this specific situation. The original Nopol patch simplifies the \emph{if} condition to (\texttt{dec == null \&\& exp == null}) and relies on checks available in the called method (\texttt{createLong(String val)}), which will return a \emph{NumberFormatException} if the format of input \emph{val} is illegal. Note the deleted predicate \texttt{(numeric.charAt(0)=='-' \&\& isDigits(numeric.substring(1)) || isDigits(numeric))} is used to check whether the variable \emph{numeric} is a legal format of number, and a \emph{NumberFormatException} will be thrown if not. Consequently, for the specific input \emph{L} and other inputs which are not legal forms of number, the desired \emph{NumberFormatException} will also be thrown after the condition is simplified. Among the 30 seeds, running Nopol+UnsatGuided with 27 seeds will get the same patch as the original Nopol run. For the other 3 seeds, running Nopol+UnsatGuided will all get the patch which adds the precondition \texttt{if(1 < val.length())} before the \emph{if} condition. After adding this precondition, the \emph{if} condition is executed only when the length of the string is larger than 1. If this precondition is not respected, the program throws the expected exception. Thus, both the original Nopol patch and the new patch generated by running Nopol+UnsatGuided are semantically equivalent to the human patch. 

\begin{figure}[t]
\centering
\begin{lstlisting}
// MANUAL PATCH
// if (val.length() == 1 && !Character.isDigit(val.charAt(0))) {
//   throw new NumberFormatException(val + " is not a valid number.");
// }
String numeric=val.substring(0, val.length()-1);
   ...
 switch (lastChar) {
   case 'L' :
      if (dec == null && exp == null && (numeric.charAt(0) == '-' && isDigits(numeric.substring(1)) || isDigits(numeric))) {
          try {
             return createLong(numeric);
           } catch (NumberFormatException nfe) { }
          return createBigInteger(numeric);
       }
      throw new NumberFormatException(val + "is not a valid number.");
    case 'f' : ...
\end{lstlisting}
 \caption{
Code snippet of buggy program version Lang\_44.
}
\label{fig:case1}
\end{figure}

For Lang\_55, the bug arises for a utility class for timing (see \autoref{fig:case2}). As discussed in \cite{defects4j-repair}, the bug appears when the user stops a suspended timer and if so, the stop time saved by the suspend action is
overwritten by the stop action. To fix the bug, the assignment of the variable \emph{stopTime} should be executed only when the state of the timer is running. The human patch adds a precondition which checks whether the state of the timer is running. The original Nopol patch and the patch generated by running Nopol+UnsatGuided (running the 30 seeds all get the same patch) both also add preconditions. 
Note the method \texttt{stop()} should be executed only when the state of the timer is suspended or running (see the if condition inside the method \texttt{stop()}), otherwise an exception will be thrown. Thus, the precondition \texttt{if (this.runningState!= STATE\_SUSPENDED)} obtained by running Nopol means the state of the timer is running. Meanwhile, given the two possible states--suspended or running, the precondition \texttt{if (this.stopTime <= this.startTime)} obtained by running Nopol+UnsatGuided can only be true when the state of the timer is running according to the logic of the utility class. Consequently, both of the two added preconditions are semantically equivalent to the precondition added by human beings.

\begin{figure}
\centering
\begin{lstlisting}
public void stop() {
        if(this.runningState != STATE_RUNNING && this.runningState != STATE_SUSPENDED) {
            throw new IllegalStateException("Stopwatch is not running. ");
        }
// MANUAL PATCH:
// if (this.runningState == STATE_RUNNING)
// NOPOL PATCH:
// if (this.runningState!= STATE_SUSPENDED) 
// NOPOL+UnsatGuided PATCH:
// if(this.stopTime <= this.startTime)
            stopTime = System.currentTimeMillis();
        this.runningState = STATE_STOPPED;
}
\end{lstlisting}
 \caption{
Code snippet of buggy program version Lang\_55.
}
\label{fig:case2}
\end{figure}

In summary, the correct patches generated by Nopol are still correct for all seeds after running Nopol+UnsatGuided. 

\textbf{Can an overfitting patch be changed into a correct one?}

It has already been shown that running Nopol+UnsatGuided can significantly change the original Nopol patch and can effectively alleviate the overfitting issue of regression introduction in the original Nopol patch. We want to further explore whether an overfitting patch can be changed into a correct one after running Nopol+UnsatGuided. Comparing the two \emph{correctness} columns under the column \emph{Nopol} and column \emph{Nopol}+\emph{UnsatGuided}, we can see that there exists one buggy version (Math\_85) for which the original Nopol patch is incorrect but the sampled patch generated by running Nopol+UnsatGuided is correct.

For Math\_85, the bug arises as the value of a condition is not handled appropriately (see \autoref{fig:case3}). The human patch changes the binary relational operator from ``>='' to ``>'', i.e., replacing \texttt{if (fa * fb >= 0.0)} with \texttt{if (fa * fb > 0.0)}. The original Nopol patch adds a precondition \texttt{if (fa * fb < 0.0)} before the \emph{if} condition in the code, which in turn will result in a self-contradictory condition and is thus incorrect. The sampled Nopol+UnsatGuided patch is adding a precondition \texttt{if (fa * fb != 0.0)} before the \emph{if} condition, which equates to the human patch semantically and is thus correct. After further checking the results for this buggy version across all 30 seeds, we find that the generated Nopol+UnsatGuided patch is the same as this patch for 21 seeds. This example shows that UnsatGuided can possibly change an original overfitting Nopol patch into a correct one. 

\begin{figure}
\centering
\begin{lstlisting}
if (fa * fb >= 0.0 ) {
        throw new ConvergenceException(
           ...
        );
   }
\end{lstlisting}
 \caption{
Code snippet of buggy program version Math\_85.
}
\label{fig:case3} 
\end{figure}

\begin{mdframed}
\textbf{Answer for RQ3}: UnsatGuided does not break any already correct Nopol patch. Furthermore, UnsatGuided can change an overfitting Nopol patch into a correct one. This is in line with our analysis of the impact of UnsatGuided on patch correctness. 
\end{mdframed}

\subsection{\textbf{RQ4}: Handling of Bug-exposing Tests} 
\label{sec:answer-rq4}

As we have seen in \autoref{sec:analysis-unsatguided}, the major challenge of using automatic test generation in the context of repair is the handling of bug-exposing tests. However, bug-exposing tests are not always generated.
Now we concentrate on the 17 buggy program versions which contain at least one bug-exposing test, i.e., rows in \autoref{tab:nopol-results} with the value of \emph{\#Bug-expo} larger than 0.

For 4 bugs (Chart\_5, Lang\_44, Lang\_51, Lang\_63), UnsatGuided works perfectly because it removes all bug-exposing tests.
Let us now explain what happens in those cases.
The column \emph{incomplete fix (\#failing)} shows that for these 4 buggy versions, the original Nopol patch does not fail on any of the bug-exposing tests, which implies that the
initial repair constraint established using the manually written test suite is strong and is likely to have reflected the desired behaviors for input points within $I_{bug}$ well.
In this case, the additional repair constraints enforced by the bug-exposing tests
have contradictions with the initial repair constraint and UnsatGuided indeed removes them, as it is designed for. 
If we do not take care of this situation and directly use all
of the automatically generated tests without any removal technique, we are
likely to lose the correct repair constraint and the acceptable patch with no overfitting issue of incomplete fixing.

For the other 13 buggy program versions, the bug-exposing tests are either not removed at all (11 cases) or partially removed (2 cases, Math\_50 and Math\_80).
The column \emph{incomplete fix (\#failing)} shows that for
these 13 buggy versions, the original Nopol patch already fails on some of the bug-exposing tests, which implies that the
initial repair constraint established using the manually written test suite does not fully reflect the desired behaviors for input points within $I_{bug}$.
Consequently, no contradiction happens during the synthesis process and these bug-exposing tests are not recognized and kept.
Now, recall that we have explained in \autoref{sec:analysis-unsatguided} that the presence of remaining bug-exposing tests does not necessarily mean worsened overfitting issue of incomplete fixing. Interestingly, this can be shown in our evaluation: 
for 9 bugs, the overfitting issue of incomplete fixing remains the same;
for 2 bugs (Math\_50 and Time\_4), the overfitting issue of incomplete fixing is reduced (the digit value in column \emph{fix completeness change (Avg\#Removedinc)} is larger than 0); and for 2 other bugs (Math\_80 and Math\_87), the overfitting issue of incomplete fixing is worsened (the digit value in column \emph{fix completeness change (Avg\#Removedinc)} is smaller than 0). To sum up, 
the unremoved bug-exposing tests do not worsen overfitting issue of incomplete fixing for the original Nopol patch in the majority of cases (11/13 cases).

Finally, let us check whether the presence of kept bug-exposing tests will have an impact on the capability of UnsatGuided in alleviating overfitting issue of regression introduction.
For the 13 buggy program versions with at least one remaining bug-exposing test, we see that UnsatGuided is still able to alleviate overfitting issue of regressions introduction.
This is the case for 5 bug versions: Math\_50, Math\_81, Math\_87, Math\_105, and Time\_4. 
This result confirms our qualitative analysis, i.e., the unremoved bug-exposing tests will not impact the effectiveness of UnsatGuided in alleviating overfitting issue of regression introduction.

\begin{mdframed}
\textbf{Answer for RQ4}: 
When bug-exposing tests are generated, UnsatGuided does not suffer from a drop in effectiveness: the overfitting issue of incomplete fixing is not
worsened in the majority of cases, and the capability of alleviating overfitting issue of regression introduction is kept. 
\end{mdframed}

\subsection{\textbf{RQ5}: Time Overhead}

The time cost of an automatic program repair technique should be manageable for being used in industry. We now discuss the time overhead incurred by UnsatGuided. 

To see the time overhead incurred, we compare the \emph{Time} column under the column \emph{Nopol} with the \emph{Avg\#Time} column under the column \emph{Nopol}+\emph{UnsatGuided}. First, we see that the approach UnsatGuided incurs some time overhead. Compared with the original repair time used by Nopol to find a patch, the average time used by running Nopol+UnsatGuided to get the patch is much longer. Second, the time overhead incurred is acceptable in many cases. Among the 42 buggy versions that can initially be repaired by Nopol, the average repair time used by running Nopol+UnsatGuided to get the patch is less than or equal to 1 hour for 28 buggy versions, which is arguably acceptable. Finally, we observe that the time overhead incurred can be extremely large sometimes. For 3 buggy versions (Chart\_26, Math\_24, and Math\_33), running Nopol+UnsatGuided will cost more than 10 hours to get the patch on average. In particular, the average time used by running Nopol+UnsatGuided to get the patch for Math\_24 is 24.1 hours. The synthesis process of Nopol is slow for those cases and the synthesis process is invoked for each generated test as required by UnsatGuided, thus the large amount of time cost is imaginable. To reduce the time overhead, future work will explore advanced patch analysis to quickly discard useless tests and identify generated tests that have the potential to improve the patch.

\begin{mdframed}
\textbf{Answer for RQ5}: UnsatGuided incurs a time overhead even though the overhead is arguably acceptable in many cases. To reduce the time overhead, more advanced techniques can be employed to analyze the automatically generated tests and discard useless ones. \end{mdframed}

\subsection{Threats to Validity}
We use 224 faults of 4 java programs from Defects4J in this study and one threat to external validity is whether our results will hold for other benchmarks. However, Defects4J is the most recent and comprehensive dataset of java bugs currently available, and is developed with the aim of providing real bugs to enable reproducible studies in software testing research. Besides, Defects4J has been extensively used as the evaluation subjects by recent research work in software testing (\cite{FLISSTA2016,FLICSE2017,FLASE2016}), and in particular by work in automated program repair (\cite{defects4j-repair,xiong2016precise}). Another threat to external validity is that we evaluate the approach UnsatGuided by viewing Nopol as the representative for synthesis-based repair techniques, and doubts may arise whether the results will generalize to other synthesis-based repair techniques. Nopol, however, is the only open-source synthesis-based repair technique that targets modern java code and can effectively repair real-life faults in real-world programs. A final threat to external validity is that only one automatic test case generation tool, i.e., EvoSuite, is used in the study. But EvoSuite is the state-of-art open source java unit test case generation tool and can target a specific java class as required by the proposed approach. Moreover, we run EvoSuite 30 times with different random seeds to account for the randomness of EvoSuite. Overall, the evaluation results are in line with our analysis of the effectiveness of UnsatGuided in alleviating different kinds of overfitting issues, and we believe the results can be generalized.

A potential threat to internal validity is that we manually check the generated patches to investigate the impact of UnsatGuided on patch correctness. We used the human patch as the correctness baseline and the human patch is also used to help us understand the root cause of the bug. This process may introduce errors. To reduce this threat as much as possible, the results reported in this paper are checked and confirmed by two authors of the paper. In addition, the whole artifact related to this paper is made available online to let readers gain a more deep understanding of our study and analysis. 

\section{Conclusion}

Much progress has been made in the area of test suite based program repair over the recent years.
However, test suite based repair techniques suffer from the overfitting problem. 
In this paper, we deeply analyze the overfitting problem in program repair and identify two kinds of overfitting issues: incomplete fixing and regression introduction. We further define three kinds of overfitting patches based on the overfitting issues that a patch has. These characterizations of overfitting will help the community to better understand and design techniques to defeat the overfitting problem in program repair. We also propose an approach called UnsatGuided, which aims to alleviate the overfitting problem for synthesis-based repair techniques. The approach uses additional automatically generated tests to strengthen the repair constraint used by synthesis-based repair techniques. We analyze the effectiveness of UnsatGuided with respect to alleviating different kinds of overfitting issues. The general usefulness of automatic test case generation in alleviating overfitting problem is also discussed. An evaluation on the 224 bugs of the Defects4J repository has confirmed our analysis and shows that UnsatGuided is effective in alleviating overfitting issue of regression introduction.

\newpage

\bibliographystyle{spbasic}
\balance
\bibliography{references} 

\end{document}